\documentclass{aa}
\usepackage{epsfig,graphics}
\usepackage[dvips]{rotating}
\usepackage{latexsym}
\usepackage{graphicx}
\usepackage{subfigure}
\usepackage{amssymb}

\newcommand{\cd}{d$^{-1}$}
\newcommand{\kms}{kms$^{-1}$}
\newcommand{\degree}{$^{\circ}$}
\newcommand{\vsini}{$v\sin{i}$}
\newcommand{\teff}{\ensuremath{T_{\rm eff}}}             
\newcommand{\logg}{\ensuremath{\log g}}                     
\newcommand{\alphaw}{$\alpha_W$}
\begin{document}
\title{A new method for the spectroscopic identification of stellar non-radial pulsation modes}
\subtitle{II. Mode identification of the $\delta$ Scuti star FG Virginis\thanks{Based on observations partly collected at The McDonald Observatory of The University of Texas at Austin and at ESO-La Silla, Proposal 69.D-0276.}}

\author{W.\ Zima \inst{1}
\and
D.\ Wright \inst{2} 
\and
J. \ Bentley \inst{2} 
\and
P. L.\ Cottrell \inst{2}
\and
U.\ Heiter \inst{3}
\and
P. Mathias \inst{4}
\and
E.\ Poretti \inst{5}
\and
H.\ Lehmann \inst{6}
\and
T.~J.\ Montemayor \inst{7}
\and
M.\ Breger \inst{1}
}

\offprints{zima@ster.kuleuven.be}

\institute
{ Institut f\"ur Astronomie der Universit\"at Wien, 
T\"urkenschanzstrasse 17, A - 1180 Vienna, Austria.  
\and
  Dept. of Physics and Astronomy, Univ. of Canterbury, Private Bag 4800, Christchurch, New Zealand
\and
Department of Astronomy and Space Physics, Uppsala University, Box 515, SE-75120 Uppsala, Sweden
\and
Observatoire de la Cote d'Azur, Departement GEMINI - UMR 6203, F-06304 Nice, France
\and
INAF - Osservatorio Astronomico di Brera, Via E. Bianchi 46, 23807 Merate(LC), Italy
\and
 Th\"uringer Landessternwarte, D-07778 Tautenburg, Germany
\and
McDonald Observatory, University of Texas at Austin, 1 University Station, C1402 Austin, Texas
}

\date{Received / Accepted }
\abstract{}
{We present a mode identification based on new high-resolution time-series spectra of the non-radially pulsating $\delta$ Scuti star FG~Vir (HD 106384, V = 6.57, A5V). From 2002 February to June a global Delta Scuti Network (DSN) campaign, utilizing high-resolution spectroscopy and simultaneous photometry has been conducted for FG~Vir in order to provide a theoretical pulsation model. In this campaign we have acquired 969 Echelle spectra covering 147 hours at six observatories.}
{The mode identification was carried out by analyzing line profile variations by means of the Fourier parameter fit method, where the observational Fourier parameters across the line are fitted with theoretical values. This method is especially well suited for determining the azimuthal order $m$ of non-radial pulsation modes and thus complementary with the method of Daszynska-Daszkiewicz (2002) which does best at identifying the degree $\ell$.}
{15 frequencies between 9.2 and 33.5~\cd~were detected spectroscopically. We determined the azimuthal order $m$ of 12 modes and constrained their harmonic degree $\ell$. Only modes of low degree ($\ell \le 4$) were detected, most of them having axisymmetric character mainly due to the relatively low projected rotational velocity of FG Vir. The detected non-axisymmetric modes have azimuthal orders between -2 and 1. We derived an inclination of 19\degree, which implies an equatorial rotational rate of 66~\kms.}
{}


\maketitle

\markboth{Zima et al.: 
Mode identification of the $\delta$ Scuti star FG Vir}
{Zima et al.: 
Mode identification of the $\delta$ Scuti star FG Vir}

\section {Introduction}
Asteroseismology permits the probing of the theory of the interior structure and evolution of oscillating stars from the precise measurement of their light variations. $\delta$ Scuti stars are particularly useful for this technique due to their large number of simultaneously excited pulsation modes having periods of a few hours. For these stars, every single detected pulsation frequency provides new independent physical information, allowing us to distinguish between different theoretical models.

The techniques of asteroseismology rely on a detailed comparison between the observed and modeled pulsation frequencies. In order to reduce the number of free parameters, we need to identify the pulsational quantum numbers associated with each mode for as many frequencies as possible. High resolution time-series spectroscopy of absorption lines is a very effective tool for mode identification and permits not only the typing of $\ell$ and $m$, but can also set constraints to the stellar inclination, the intrinsic pulsation amplitude and involved temperature variations.

A key object for successful asteroseismic studies is FG Vir, one of the best studied $\delta$ Scuti stars. It has been the target of numerous photometric and spectroscopic observing campaigns during the past 20 years.  Despite intense efforts no definitive pulsational modeling has been achieved yet, although we already have detailed knowledge about its rich frequency spectrum from the observational side. This is mainly due to a lack of good identifications of its pulsation modes. The determination of $\ell$ and $m$ for at least a few modes would decrease the number of degrees of freedom of possible theoretical pulsation models.

We applied the Fourier parameter fit (FPF) method (Zima 2006), hereafter referred to as Paper I, to spectra of FG Vir for the identification of the detected pulsation modes. This method is ideally suited for narrow lined stars pulsating with low-degree modes and enables the determination of $\ell$ and $m$, as well as other pulsational parameters. 

This paper is constructed in the following way. First, we provide a short summary of the observational history of FG Vir, followed by a description of the new spectroscopic measurements and the data reduction. The spectroscopic detection of pulsation frequencies from the analysis of line profiles is discussed. Finally, we report the mode identification by means of the FPF method.

\section{Previous spectroscopic observations of FG Vir}
Eggen (1971) discovered the variability of FG Vir in 1970 during three hours of photometry detecting one period of 0.07~d (14.2~\cd)  and a semi-amplitude of 25 mmag. Subsequent photometric observations of this object (Dawson et al. 1995) revealed a rich pulsation spectrum making it a target of further extended studies in the following years. 

Pioneering work has been done by Mantegazza et al. (1994), who realized that only a combined photometric and spectroscopic approach would yield significant progress for pulsational modeling. Eight nights of single-site differential photometry in $V$ revealed seven frequencies. The high-resolution spectra obtained during a single night showed that FG Vir has a low projected rotational velocity of about \vsini=21 \kms~resulting in narrow absorption lines. The dominant mode was identified as probably the first radial overtone mode - contrary to the results of Dawson et al. (1995).

Viskum et al. (1998) applied the equivalent width method to FG Vir in order to derive $\ell$ for eight frequencies. Their results were in good agreement with the identification by Breger et al. (1999), who applied a photometric mode identification based on phase shifts and amplitude ratios of frequencies in two different pass bands. Balona (2000) showed that the two applied techniques are not independent. Furthermore, Viskum et al. (1998) claimed a rotation period of 3.5 d derived from variations in the time series which would indicate an equatorial rotation velocity of about 33 \kms~and an inclination of 40$^\circ$.

A detailed atmospheric study of FG Vir was made by Mittermayer \& Weiss (2003) in order to derive its chemical composition. Their abundance analysis showed a solar-like pattern with slight overabundances of Ba and an underabundance of C and S. A metal overabundance claimed by Russell (1995) could not be confirmed. The analysis was hampered by the pulsational broadening of the absorption lines, which required an unrealistically high macro-turbulence in order to model the spectra. 

Recently, Mantegazza \& Poretti (2002) published results from five nights of high-resolution time-series spectroscopy obtained during the 1995 DSN campaign of FG Vir. They analyzed the line profile variations with the pixel-by-pixel method and were able to detect ten previously known frequencies in the radial velocity and pixel-by-pixel variations. Their identification revealed eight axisymmetric modes ($m=0$) and only two modes with $\ell \ge 1$. They derived an inclination value of 15$^\circ$ which is much lower than the value proposed by Viskum et al. (1998), leading to a much higher equatorial rotation velocity of 80 \kms. By combining flux and velocity measurements they were able to derive the following physical parameters for the dominant mode: $\ell=1$, $m=0$, $a_s=0.00147\pm0.026$, $i=15^\circ$, $f_r=-4.3$ and $f_i=14$, where $f_r$ and $f_i$ are the real and imaginary part of $f$, the ratio of the stellar flux to radius variations.

Meanwhile, a number of extended global photometric observing campaigns, carried out by the Delta Scuti Network (Breger et al. 1998, 2004, 2005) revealed 79 pulsation and combination frequencies. The existence of several very close frequencies, some being separated less than 0.01~\cd~(Breger \& Pamyatnykh 2006), makes seismic modeling with modes of harmonic degree higher than 3 necessary. 

Daszynska-Daszkiewicz et al. (2002) developed a new method for the determination of $\ell$ and $f$ from photometric amplitudes in different filters and radial velocity amplitudes. They applied their method on the simultaneously obtained photometric and spectroscopic data of FG Vir in 2002 (Daszynska-Daszkiewicz et al. 2005). Twelve pulsation modes have been detected spectroscopically in the radial velocity curve and all of them have photometric counterparts in $v$ and $y$ (Breger et al. 2005). A comparison of the empirical and theoretical $f$ values of these modes, adopting different descriptions of the mixing length parameter $\alpha$, yielded the best agreement for $0 \le \alpha \le 0.5$, meaning that convection is relatively inefficient in FG Vir.

\section {Observations and data reduction}
\subsection{New measurements}
During 2002 February to June we carried out an observing campaign for the $\delta$ Scuti star FG Vir in the framework of the Delta Scuti Network (Rodler 2003, Zima 2002). Photometric as well as spectroscopic time-series observations were acquired in order to enable the detection and identification of more pulsation frequencies, pinpoint stellar parameters like the inclination and the equatorial rotational velocity, and consequently improve stellar pulsation models. The combination of simultaneous photometry and spectroscopy enabled mode identification methods, which make use of flux and velocity variations and thus provide a test for the efficiency of convection (Daszynska-Daszkiewicz et al. 2005).

To avoid aliasing effects caused by observing gaps during daylight and unpassable weather, multi-site observations from different continents were necessary. Daily aliasing, which creates spurious peaks in a Fourier spectrum spaced by 1~\cd, can seriously prevent the detection of intrinsic signal. Since FG Vir is located near the celestial equator ($\delta=-5^\circ$) observations from both hemispheres were possible. We selected telescopes of the 2-meter class at sites in Germany, France, USA, Chile, South Africa and New Zealand which permitted data acquisition of high-dispersion (R $\ge$ 30000) spectra and of high signal-to-noise (S/N $\ge$ 100) covering a large wavelength range. The chosen integration times are between 7 and 10 minutes, which is about 20 \% of the shortest detected period of FG Vir and 6 \% of the dominant frequency at 12.716~\cd. The observations at ESO were made with a shorter integration time of 5 minutes to explore the higher frequency domain.

Spectroscopy of FG Vir was carried out at the following six sites:
\begin{enumerate}
\item Measurements at the Mt. John University Observatory (New Zealand) were carried out with the High Efficiency and Resolution Canterbury University Large Echelle Spectrograph (HERCULES) attached to the 1m telescope. The wavelength range was 4585-5837~\AA~and the resolution R=35000. The integration time was 7 minutes. The observers were Duncan Wright, John Bentley and Peter Cottrell.
\item The South African Astronomical Observatory (SAAO, South Africa) data were acquired with the GIRAFFE spectrograph attached to the 1.9m telescope. The wavelength range was 4400 to 7000~\AA~at a resolution of $\sim 32000$. The integration time was 7 minutes. The observer was Wolfgang Zima.
\item The observations at the 2.1m telescope of the McDonald Observatory (Texas, USA) were made with the Sandiford Cassegrain Echelle Spectrograph at a resolution of R=60000 in a wavelength range of 4595-5194~\AA. The integration time was 7 minutes. The observer was Ulrike Heiter.
\item The measurements at Tautenburg Observatorium (Germany) were obtained at the 2m telescope equipped with a Coude Echelle spectrograph with a resolution of R=63000 and in a wavelength range of 4708-7362~\AA. The integration time was 7 minutes. The observer was Holger Lehmann.
\item At the European Southern Observatory (ESO, Chile) the 2.2m telescope was used with the Fiber-fed Extended Range Optical Spectrograph (FEROS). The wavelength range was 4379-6549~\AA~at a resolution of R=48000 and an integration time of 5 minutes. The observer was Ennio Poretti.
\item Observations at the Observatoire de Haute-Provence (OHP, France) were obtained using the AURELIE spectrograph attached to the 1.52m telescope. The spectral domain covers the range 4470-4540~\AA~with a resolution power R=55000. The integration time was typically 10 minutes. The observer was Philippe Mathias.
\end{enumerate}

A journal of the observations is given in Table~\ref{tab:observations}.
\begin{table}[!ht]
\centering
\caption{List of spectroscopic observations during the 2002 FG Vir campaign.}
\begin{tabular}{c|c|c|c|c} \hline
\multicolumn{2}{c}{HJD -2450000} & Length & No. of & Obs.\\
Begin & End & h & spectra & \\\hline
2295.3268 & 2295.3855 & 1.41 & 3 & OHP\\
2314.3641 & 2314.4751 & 2.66 & 16 & OHP\\
2315.3459 & 2315.4718 & 3.02 & 18 & OHP\\
2316.3994 & 2316.4349 & 0.85 & 6 & OHP\\
2318.3518 & 2318.4769 & 3.00 & 18 & OHP\\
2320.3481 & 2320.4708 & 2.94 & 18 & OHP\\
2322.3591 & 2322.4786 & 2.87 & 17 & OHP\\
2324.3364 & 2324.4596 & 2.96 & 18 & OHP\\
2326.3366 & 2326.4614 & 2.99 & 18 & OHP\\
2327.3375 & 2327.3506 & 0.31 & 3 & OHP\\
2328.3371 & 2328.4481 & 2.66 & 15 & OHP\\
2333.9978 & 2234.1754 & 4.26 & 22 & Mt. John\\
2336.3856 & 2336.6540 & 6.44 & 42 & SAAO\\
2336.9873 & 2337.2149 & 5.46 & 26 & Mt. John\\
2337.3788 & 2337.6577 & 6.70 & 51 & SAAO\\
2339.3676 & 2339.6485 & 6.74 & 51 & SAAO\\
2342.3519 & 2342.6555 & 7.29 & 55 & SAAO\\
2343.4398 & 2343.6018 & 3.89 & 30 & SAAO\\
2345.4291 & 2345.5142 & 2.04 & 16 & SAAO\\
2355.9534 & 2356.1509 & 4.74 & 34 & Mt. John\\
2356.6687 & 2356.9986 & 7.92 & 45 & McDonald\\
2356.9504 & 2357.1241 & 4.17 & 31 & Mt. John\\
2357.6545 & 2357.8752 & 5.30 & 36 & McDonald\\
2357.9702 & 2358.1172 & 3.53 & 24 & Mt. John\\
2358.8771 & 2358.9689 & 2.20 & 16 & McDonald\\
2359.6336 & 2359.9914 & 8.59 & 50 & McDonald\\
2359.9489 & 2360.1351 & 4.47 & 29 & Mt. John\\
2360.6296 & 2360.9644 & 8.04 & 51 & McDonald\\
2361.3612 & 2361.6161 & 6.12 & 39 & Tautenburg\\
2362.3751 & 2362.6369 & 6.28 & 40 & Tautenburg\\
2392.5203 & 2392.7313 & 5.06 & 38 & ESO\\
2393.7066 & 2393.8026 & 2.30 & 19 & ESO\\
2394.6204 & 2394.7836 & 3.92 & 26 & ESO\\
2409.8842 & 2410.0305 & 3.51 & 24 & Mt. John\\
2410.8931 & 2410.9571 & 1.53 & 12 & Mt. John\\
2411.9176 & 2411.9853 & 1.63 & 12 & Mt. John\\ \hline

\end{tabular}
\label{tab:observations}
\end{table}



\subsection{Data reduction}


The data from SAAO, ESO, McDonald and OHP were reduced with standard IRAF (Image Reduction and Analysis Facility) routines. We applied bias, flatfield and stray-light corrections to the spectra. Every single order was extracted and a wavelength calibration by utilization of the ThAr comparison spectra was applied. The reduction of the Tautenburg data was carried out in the same way as described above with an additional radial velocity zero point correction using telluric lines. The data from Mt. John Observatory were reduced with a standard reduction package (Skuljan et al. 2000). 
An adjustment of the wavelength to the heliocentric frame was carried out and the times of observation were transferred to the Heliocentric Julian Date. Finally, we normalized the spectra to the pseudo continuum of each order.

\begin{figure}[!ht]
\centering
  \includegraphics*[height=80mm,bb=292 48 536 547,clip,angle=-90]{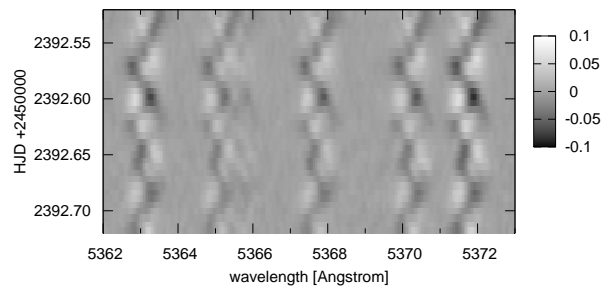}
\caption{Profile variations of the absorption lines between 5362 and 5373 \AA~during one night of observation at ESO. The residuals after subtraction of the mean profile are shown in gray scales with the scale on the right-hand side (continuum units).}
\label{fig:lpv5360b5375}
\end{figure}

An example of the observed line profile variability (LPV) of FG Vir is given in Fig.~\ref{fig:lpv5360b5375}. It shows the intensity deviation from the mean for a set of high S/N spectra in the range 5360 to 5375 \AA~taken during one observation night at ESO. The synchronous movement of the absorption lines is mainly due to the dominant mode at 12.7~\cd. 


\subsection{Selection of lines} \label{sec:seloflines}
The most important criteria for the selection of the absorption lines to be analyzed were identical temporal behavior and the avoidance of blending. 

We calculated model atmospheres and synthetic spectra using stellar parameters and abundances derived by Mittermayer \& Weiss (2003), hereafter MW03, to support the search for unblended lines. Unfortunately, we did not find a single line which met the requirements in the overlapping wavelength region of all six observatories. The range below 4700~\AA~was omitted for the analysis, due to the rapidly decreasing S/N of the SAAO data towards the blue. Thus, we decided to omit the McDonald and OHP data for the overall frequency analysis and mode identification in favor of more suitable lines. This resulted in a loss of potential frequency resolution due to the shorter time base, but enabled us to study line profiles of higher quality. A separate analysis of the LPV of the OHP spectra is reported at the end of Section~\ref{sec:period}.

\begin{figure}[!ht]
\centering
  \includegraphics*[bb=554 692 53 22,height=90mm,clip,angle=90]{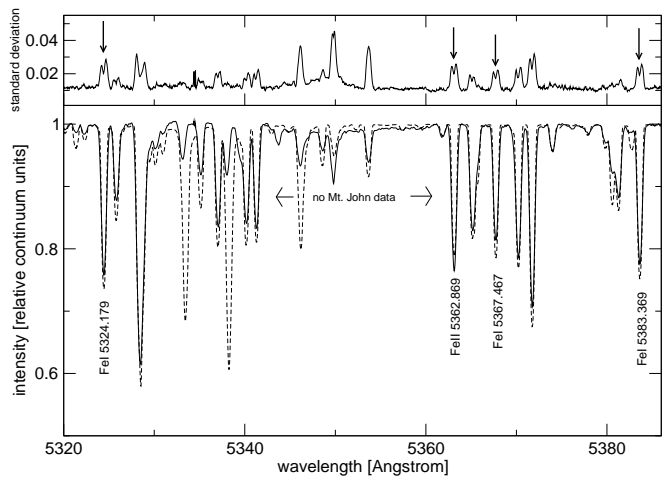}
\caption{Spectrum between 5320 and 5400~\AA. The lower panel shows the mean spectrum (solid line) of the four selected observatories and lines, which were chosen for a closer analysis are marked. A synthetic spectrum computed from a static model atmosphere (MW03) is shown as dotted line. The standard deviation from the mean is plotted in the top panel.}
 \label{fig:range5300}
\end{figure}

We selected four unblended absorption lines shown in Fig.~\ref{fig:range5300} for the mode identification. In the lower panel the mean spectrum (solid line) is plotted together with a synthetic spectrum (dotted line) calculated from a static model atmosphere\footnote{The synthetic spectrum was computed with the program SYNTH (Piskunov 1992) using parameters of the spectral lines from VALD (Piskunov et al. 1995, Kupka et al. 1999, Ryabchikova et al. 1999). The model atmosphere was calculated using the ATLAS9 code (Kurucz 1993).} adopting the following input parameters: \teff~=7425~K, \logg=3.9, micro turbulence velocity $v_{\rm micro}$=3.9~\kms~and \vsini=21.3~\kms~(from MW03). The upper panel shows the standard deviation from the mean. Due to the pulsationally induced intensity variations, the deviation is larger at the position of the lines than at the continuum.

The following lines were chosen: \ion{Fe}{i}~$\lambda$5324.179~\AA, \ion{Fe}{ii}~$\lambda$5362.869~\AA, \ion{Fe}{i}~$\lambda$5367.467~\AA~and \ion{Fe}{i}~$\lambda$5383.369~\AA. The wavelength values of the lines were taken from the Vienna Atomic Line Database (VALD) and chosen as zero points in the conversion of \AA ngstrom to \kms.  The data were weighted according to their S/N measured in two continuum-regions between 5310 and 5400~\AA. The measurements taken at ESO, Mt. John and Tautenburg all have a similar average S/N of about 200, whereas the quality of the SAAO observations is much lower with about 80. 

\begin{figure*}[!ht]
\begin{center}
\begin{tabular}{cc}
  \includegraphics*[width=85mm,bb=41 52 517 760,clip,angle=0]{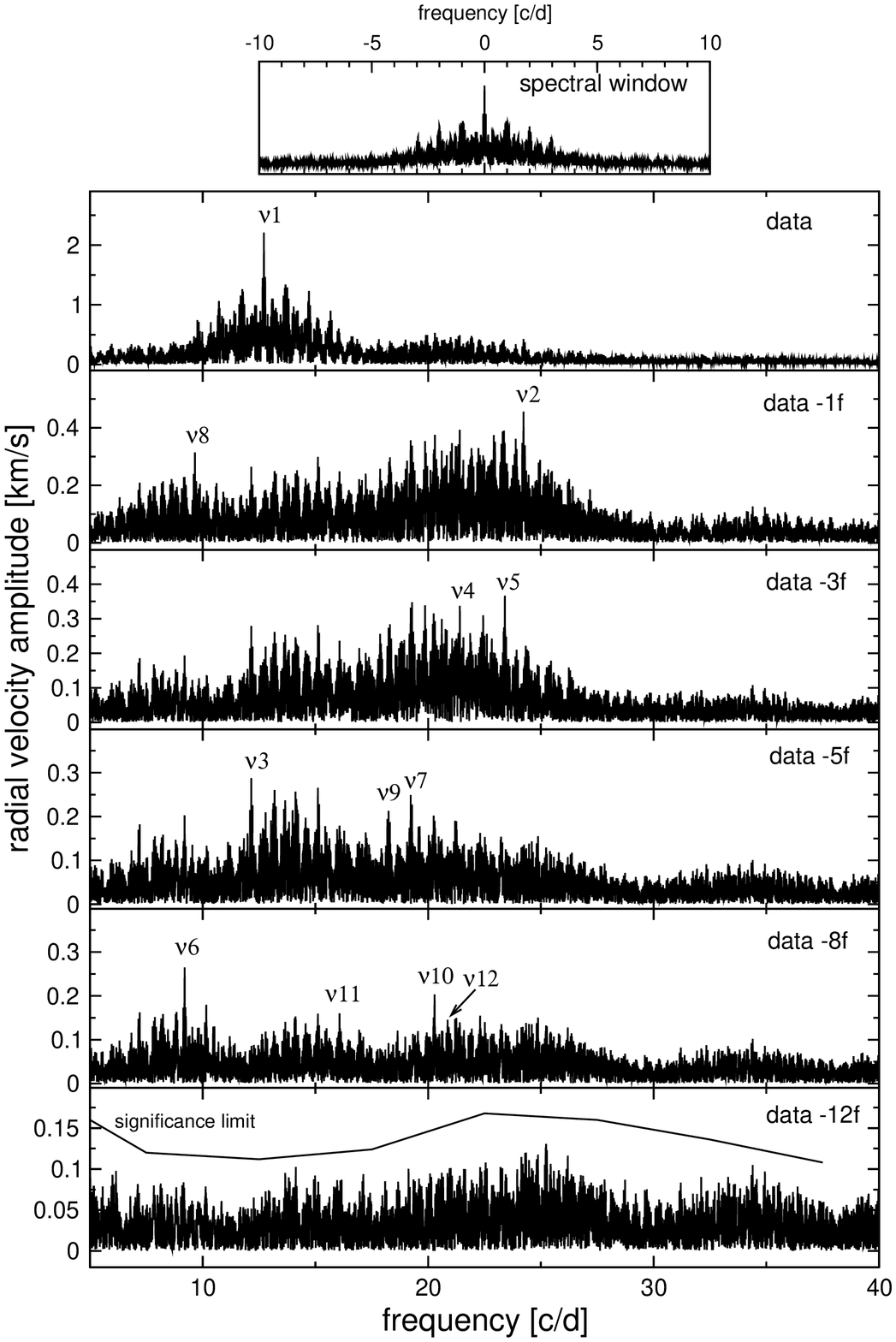} &
  \includegraphics*[width=91mm,bb=36 52 544 790,clip,angle=0]{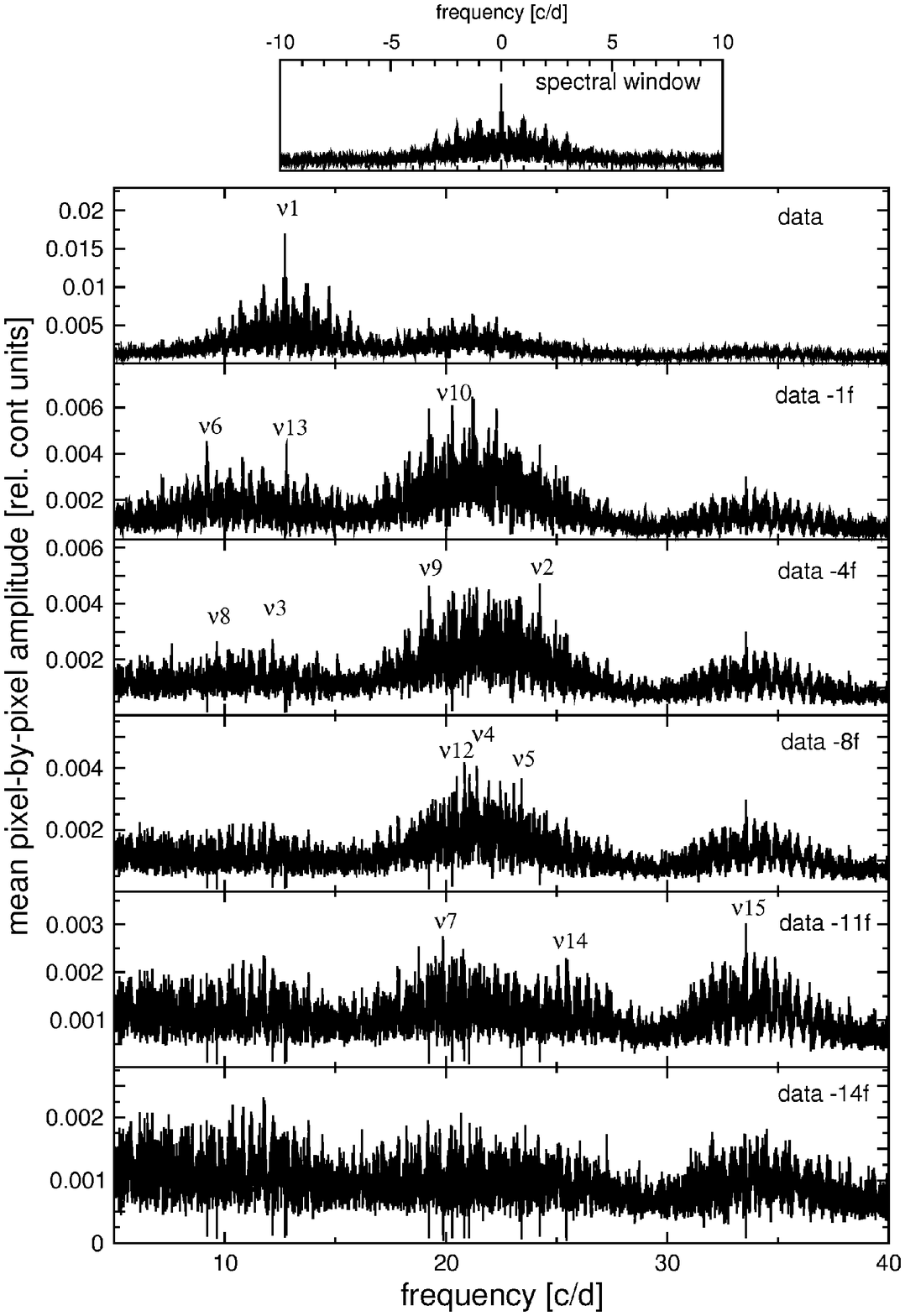}\\
\end{tabular}
\end{center}
\caption{Fourier analysis of the radial velocity variations (left panels) and the mean Fourier spectrum across the line (right panel) of the combined line profile. From top to bottom, every panel shows an amplitude spectrum prewhitened by all frequencies marked in the panels above. The significance limit in the bottom panel is calculated according to 4*noise over a 10~\cd-range. Note that the scale of the ordinate is changed in the different panels.}
 \label{fig:r5310ampfm}
\end{figure*}
We examined the temporal behavior of the four lines separately by means of a period analysis of their equivalent width variations. An identical temporal behavior is especially important when combining line profiles to a single profile since deviations would distort the combined profile. No significant differences in the radial velocity amplitudes and equivalent width variations of the four lines could be detected. Therefore, it was safe to combine them by computing a mean profile after the wavelength values of every line were converted to the Doppler scale centered on the values given by VALD. The mean S/N of this combined profile is 240, which is an appropriate value for the application of the FPF method.

Finally, the resulting combined profiles were deconvoluted with the instrumental profiles, derived for the data set of each observatory separately. Such an approach was necessary in order to eliminate different line broadenings due to the different instrumental profiles. 

\section{Period analysis} \label{sec:period}
We examined the pulsational content of the combined line profile by means of a separate analysis of the radial velocity and pixel-by-pixel variations. 

Our approach for the detection of the frequencies was carried out following the recursive procedure described below. At each step a Fourier spectrum of the original or prewhitened data was calculated. The peaks with the highest amplitudes were selected and examined for their statistical significance. From all detected frequencies, a multi-periodic least-squares fit was calculated and the data were prewhitened with the derived fit. This cycle was performed until the amplitude spectrum was void of statistically significant periodicities. For our data, we adopted the criterion that a peak is statistically significant if its S/N is above 4.0 in amplitude (Breger et al. 1993).


The left hand side of Fig.~\ref{fig:r5310ampfm} shows the Fourier analysis of the radial velocity variations of FG Vir. According to the spectral window in the top panel, the daily aliasing is low, giving the 1~\cd-aliasing peaks only 50~\% the height of the central peak. The raw Fourier spectrum is dominated by the frequency $f_1=12.716$~\cd~also showing up in the photometric data as the mode of highest amplitude. The order of the modes' radial velocity amplitudes is very similar to the photometric amplitudes. This is not really a surprise, since both values are integrated across the stellar disk and the detection of low-degree modes is favored. During each iteration of the frequency search we replaced the frequency values of the detected modes with the more accurately known values from the 2002-2004 photometry (Breger et al. 2005). Consequently, during the least-squares fit the frequency values were fixed and only the zero point, amplitude and phase were improved. 

\begin{figure*}[!ht]
\centering
  \includegraphics*[bb=50 430 510 760,width=130mm,clip,angle=0]{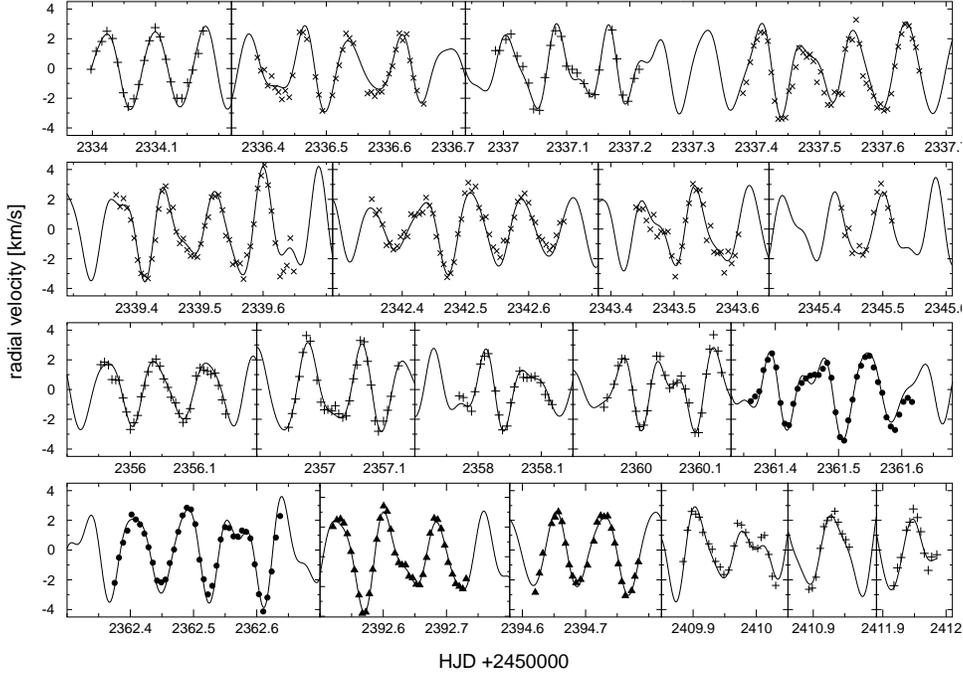}
  \caption{Observations and 12f-fit of the radial velocity variations of the averaged line profile after shift to the heliocentric frame. (+ Mt. John, $\times$ SAAO, $\bullet$ Tautenburg, $\blacktriangle$ ESO)}
 \label{fig:r5310firstmom}
\end{figure*}

In Table~\ref{tab:frequs_rv} we list all 15 statistically significant periodic terms between 9 and 34~\cd~detected in this approach. Note that in the following we assign to spectroscopically detected terms the prefix $\nu$, whereas the modes known from photometry are denoted with the prefix $f$ as indicated by Breger et al. (2005). We report the radial velocity observations together with a 12-frequency fit in Fig.~\ref{fig:r5310firstmom}. 

Eleven detected modes correspond to the strongest photometric components having $y$-amplitudes between 22 and 1~mmag. There are two modes, $f_{12}=23.397$~\cd~and $f_{10}=24.194$~\cd~having photometric $y$-amplitudes above 1 mmag, which do not appear in our frequency solution. Both modes are components of close frequency pairs of which one component has been detected spectroscopically. The frequency resolution of the spectroscopic data of~0.019~\cd~does not permit a separation of the first pair, but the separation of the two close modes at 24.2~\cd~should be sufficient. Indeed, the highest remaining peak of the residuals after subtraction of the 12 modes is at 25.21~\cd, which turns out to be an alias of $f_{10}$. The spectroscopic detection of $\nu_{12}=20.83$~\cd, which has a photometric $y$-amplitude of only 0.33 mmag is remarkable. No harmonics or combination frequencies could be found. 
\begin{figure}[!ht]
\begin{center}
\begin{tabular}{c}
  \includegraphics*[height=72mm,bb=78 3 579 725,clip,angle=-90]{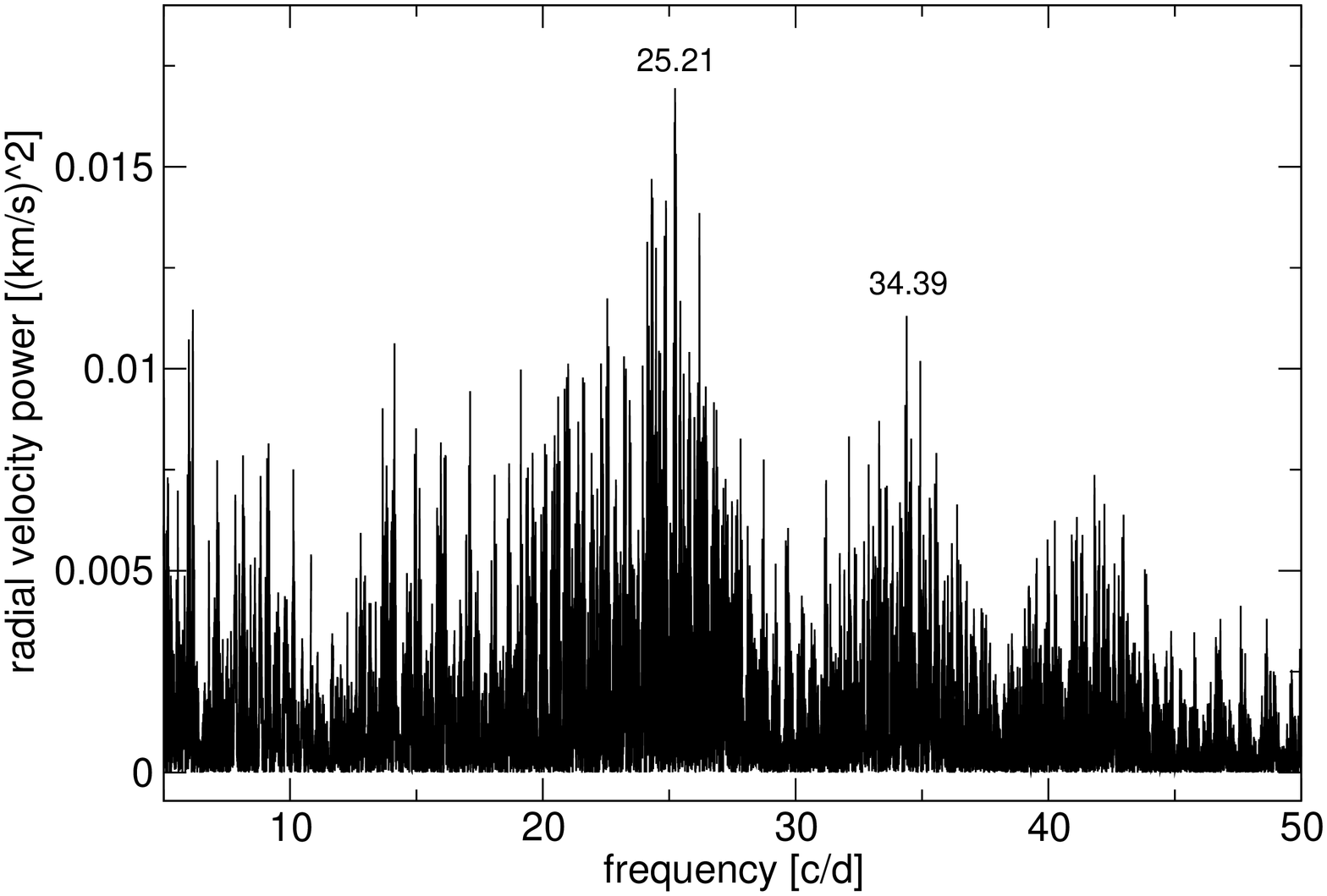} \\
  \includegraphics*[height=72mm,bb=78 3 579 725,clip,angle=-90]{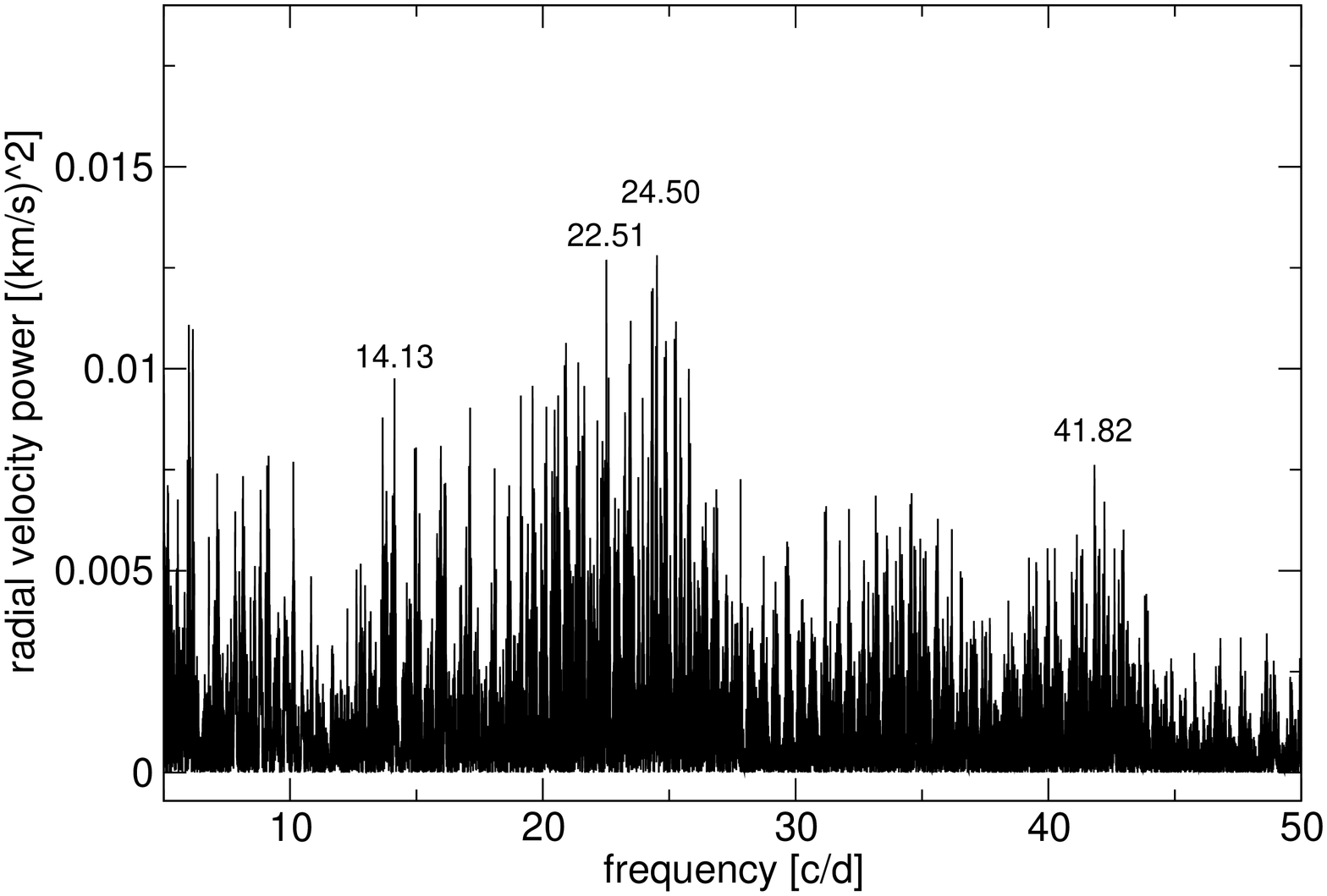}\\
\end{tabular}
\end{center}
  \caption{Residual power of the radial velocity between 5 and 50~\cd. Left panel: The residuals after prewhitening 12 frequencies. Two of the highest remaining peaks are identified with modes known from photometry (the term at 25.2~\cd~is a one-day alias of $f_{10}$). Right panel: After additional prewhitening of these two frequencies, none of the remaining peaks can be identified with a photometric counterpart. Especially in the regions around 14~\cd, from 20 to 28~\cd, around 35~\cd~and 43~\cd~the noise level is noticeably increased. A similar situation was observed in the photoelectric data from 1995 (Breger et al. 1998).}
 \label{fig:r5310residuals}
\end{figure}

We compared the highest remaining peaks, which are all below the significance limit, after prewhitening of all detected modes with the frequencies detected by Breger et al. (2005). The residual power spectra are displayed in Fig.~\ref{fig:r5310residuals}. The only components appearing also in the 79 photometric frequencies are $f_{10}$ and $f_{55}=34.39$~\cd~at very low S/N. Other residual peaks are apparently not related to detected photometric frequencies. There remains an increased noise level in some frequency domains, especially between 20 and 30~\cd~where photometry has revealed many additional frequencies. 

\begin{table}[!ht]
\centering
\caption{Frequencies detected by Fourier analysis of the radial velocity (RV) and pixel-by-pixel (PbP) variations. The frequency values are taken from photometry (Breger et al. 2005). The amplitudes are calculated from a multi-periodic least-squares fit to the data. The PbP amplitude is the integral of the amplitude across the line. The photometric amplitudes are derived from a 60 frequency least-squares fit to the 2002 data.}
\footnotesize
\begin{tabular}{l|c|c|c|c|c} \hline
id & freq. & RV amp. & PbP amp. & $y$ amp. &$v$ amp.\\
 & \cd & \kms & \kms &mmag & mmag\\
 &  & $\pm$0.03 & $\pm$0.02 & $\pm$0.07 & $\pm$0.08\\ \hline
$\nu_6$   &      9.199    & 0.28 & 0.33 & 2.73 & 3.80\\
$\nu_8$   &      9.656    & 0.28 & 0.20 & 3.62 & 5.14\\
$\nu_3$   &      12.154   & 0.36 & 0.21 & 4.10 & 6.08\\
$\nu_1$   &      12.716   & 2.06 & 1.16 & 22.10 & 32.13\\
$\nu_{13}$  &      12.794   & - & 0.29 & 0.68 & 0.82\\ 
$\nu_{11}$  &      16.071   & 0.16 &-&1.23&1.61\\
$\nu_9$  &      19.227   & 0.18  & 0.29 & 1.73 & 2.52\\
$\nu_7$  &      19.867   & 0.31  & 0.20 & 1.96 & 2.79\\
$\nu_{10}$  &      20.287   & 0.23 & 0.35 & 1.18 & 1.84\\
$\nu_{12}$  &      20.834   & 0.15 & 0.27& 0.33& 0.49\\ 
$\nu_4$   &      21.051   & 0.38 & 0.23 & 3.07 & 4.32\\
$\nu_5$   &      23.403   & 0.36 & 0.25 & 3.98 & 5.33\\
$\nu_2$   &      24.227   & 0.43 & 0.25 & 4.26 & 5.65\\
$\nu_{14}$  &      25.433   & - & 0.16 & 0.98&1.34\\
$\nu_{15}$  &      33.551   & - & 0.18 &-&-\\ \hline
\end{tabular}
\label{tab:frequs_rv}
\end{table}
The frequency analysis of the pixel-by-pixel variations shows a slightly different picture. The frequency spectra at subsequent stages of prewhitening are reported in the right hand side of Fig.~\ref{fig:r5310ampfm}. Here, the mean of every Fourier spectrum computed for each wavelength bin across the profile is displayed. Since here, in contrast to the Fourier analysis of the radial velocity, the phase information across the line is neglected, high degree modes can be detected.

The frequency spectrum is again dominated by $\nu_1$. The subsequent modes have relatively low radial velocity amplitudes. The term $\nu_{13}=12.79$~\cd, which was not detected in the radial velocity variations, shows up here as the fourth highest peak. Intriguing is the detection of the combination mode $\nu_{15}=33.55$~\cd$=\nu_{1}+\nu_{12}$. This is the only term not found among the 79 photometrically detected frequencies and it is the only spectroscopically revealed combination frequency. After prewhitening the data for 14 frequencies, none of the remaining peaks can be identified with a photometric counterpart. 

We determined the mean radial velocity of FG Vir relative to the heliocentric frame by cross correlating the mean of the observed spectrum with a synthetic spectrum between 5310 and 5390~\AA. Using this method the velocity information of all lines in that range can be utilized. The location of the main peak in the cross correlation function provides the shift between these two spectra. Thereby we obtained a value of +14.4$\pm$0.2~\kms.

The measurements at ESO have a short integration time of 5 minutes reducing the effects of phase smearing for higher frequencies. We carried out a separate frequency analysis of the radial velocities and pixel-by-pixel variations. No additional statistically significant frequencies above $\nu_{15}=33.55$~\cd~could be detected spectroscopically from these data. More multi-site observations having such short integration time would be required to permit an improved frequency analysis of this high frequency range.

Comparing our frequency analysis to Mantegazza \& Poretti (2002), hereafter MP02, the striking difference to their results is the radial velocity amplitude of $\nu_1$. A re-analysis of the 1995 data\footnote{Kindly provided by L. Mantegazza} has revealed some inconsistencies in the determination of the radial velocity values as obtained from the cross correlation of the normalized spectra with the average spectrum. Indeed, when calculating the radial velocity amplitudes as first moment (i.e., the line barycentre) the agreement with the values listed in Table~2 is excellent\footnote{The new radial velocity amplitudes for the 10 terms solution provided by MP02 in their Table~1 are, from top to bottom: 2.10, 0.51, 0.31, 0.53, 0.34, 0.41, 0.35, 0.34, 0.38 and 0.13~kms$^{-1}$}. Therefore, there is no reason to suspect the variability of the radial velocity amplitude of the $\nu_1$ term, also considering that its photometric amplitude is stable. Moreover, we note that the analysis of the 1992 data (Mantegazza et al. 1994) supplied very similar radial velocity amplitudes.


Interestingly, MP02 were able to detect three photometric frequencies at 34.12, 21.23 and 21.55~\cd, which do not show up in our data. An explanation might be that they used deeper absorption lines in their analysis resulting in a higher ratio of pixel-by-pixel amplitudes to the observational noise.

A separate frequency analysis of the radial velocities and line profile variations was also undertaken for the OHP spectra using the mean of the two metallic lines \ion{Ti}{ii} 4501 and \ion{Fe}{ii} 4508~\AA. Radial velocities were computed using two methods: a Gaussian fit, and the first normalized moment.

A Fourier analysis clearly shows the $\nu_1$ frequency, which has an amplitude of 2.1~\kms. After prewhitening with $\nu_1$, many peaks appear, the dominant one being around 21.2 /d (alias of $\nu_2$). Many significant peaks are detected: 12.15 ($f_3$), 9.66 ($f_8$), but they are polluted by aliasing problems due to the mono-site situation.
Table 3 lists the radial velocity amplitudes of the OHP spectra computed by a simultaneous multi-periodic least-squares fit taking the frequencies from Table 2.
\begin{table}[!ht]
\label{tab:ohp}
\begin{center}
\caption{Radial velocity amplitudes of the OHP spectra computed from a multi-periodic least-squares fit using the frequencies listed in Table 2.}
\begin{tabular}{r|l}
\hline
\multicolumn{1}{c}{frequency} &
\multicolumn{1}{c}{RV amplitude}\\
\cd & \kms\\
\hline
12.716 & 2.10\\
24.227 & 0.33\\
12.154 & 0.51\\
21.051 & 0.52\\
23.403 & 0.31\\
 9.199 & 0.18\\
19.867 & 0.13\\
 9.656 & 0.25\\
19.227 & 0.37\\
20.287 & 0.23\\
16.071 & 0.22\\
20.834 & 0.21\\
12.794 & 0.19\\
25.433 & 0.19\\
33.551 & 0.04\\
\hline
\end{tabular}
\end{center}
\end{table}

In 1999, single-site high-resolution Echelle spectroscopy of FG Vir was carried out during the nights of May 3, 6, and 7 at McDonald Observatory by T. J. Montemayor. Fourtytwo spectra between 5500 and 6000~\AA, having an average S/N of 220, were acquired. We used these data to check the stability of the radial velocity amplitude of $\nu_1$ and the mean radial velocity relative to the heliocentric frame. 

Only $\nu_1$ could be detected in the variations of the radial velocity and pixel intensities of the \ion{Fe}{i} 5525~\AA~line. Other periodicities could not be resolved due to the small number of data points. We derived a radial velocity amplitude of $1.81\pm0.2$~\kms, which is consistent with the value derived in 2002.

The mean radial velocity, +14.7$\pm$0.5~\kms, was derived from lines in the range between 5520 and 6002~\AA~by the same procedure as described above for the 2002 data. Therefore, we can say that no significant radial velocity shift occurred between the three years. 

Also considering that a new analysis of the 1992 and the 1995 data supplied mean radial velocity values of $14.4\pm 0.3$ and $15.4\pm0.2$~\kms, respectively, we can say that no significant radial velocity shift occured between 1992 and 2002.

\begin{figure*}[!ht]
\centering
\begin{center}
 \begin{tabular}{ccc}
\vspace*{-2mm} 
  \includegraphics*[width=50mm,bb=73 104 465 519,clip,angle=0]{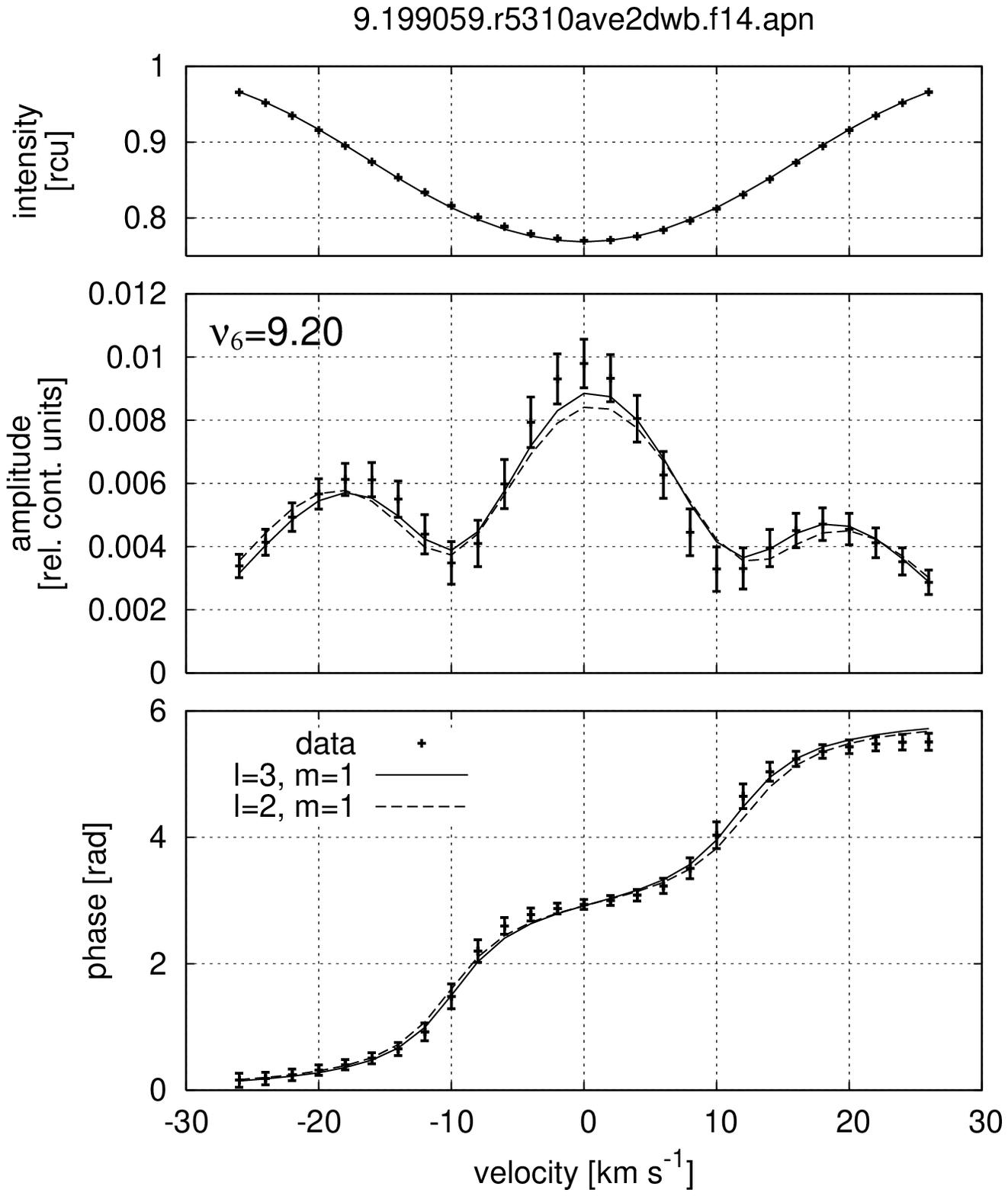} &
  \includegraphics*[width=45mm,bb=107 104 465 519,clip,angle=0]{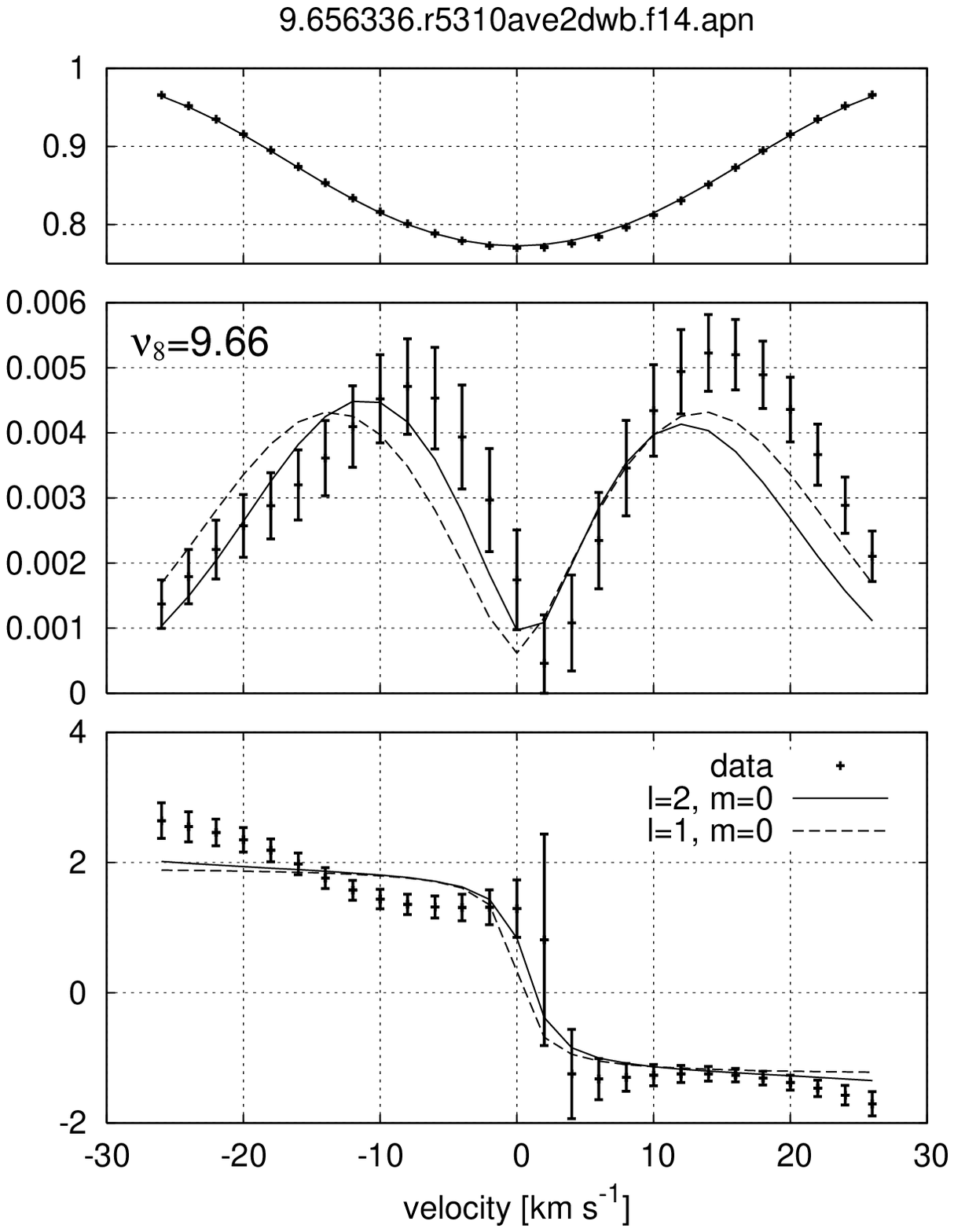}&
  \includegraphics*[width=45mm,bb=107 104 465 519,clip,angle=0]{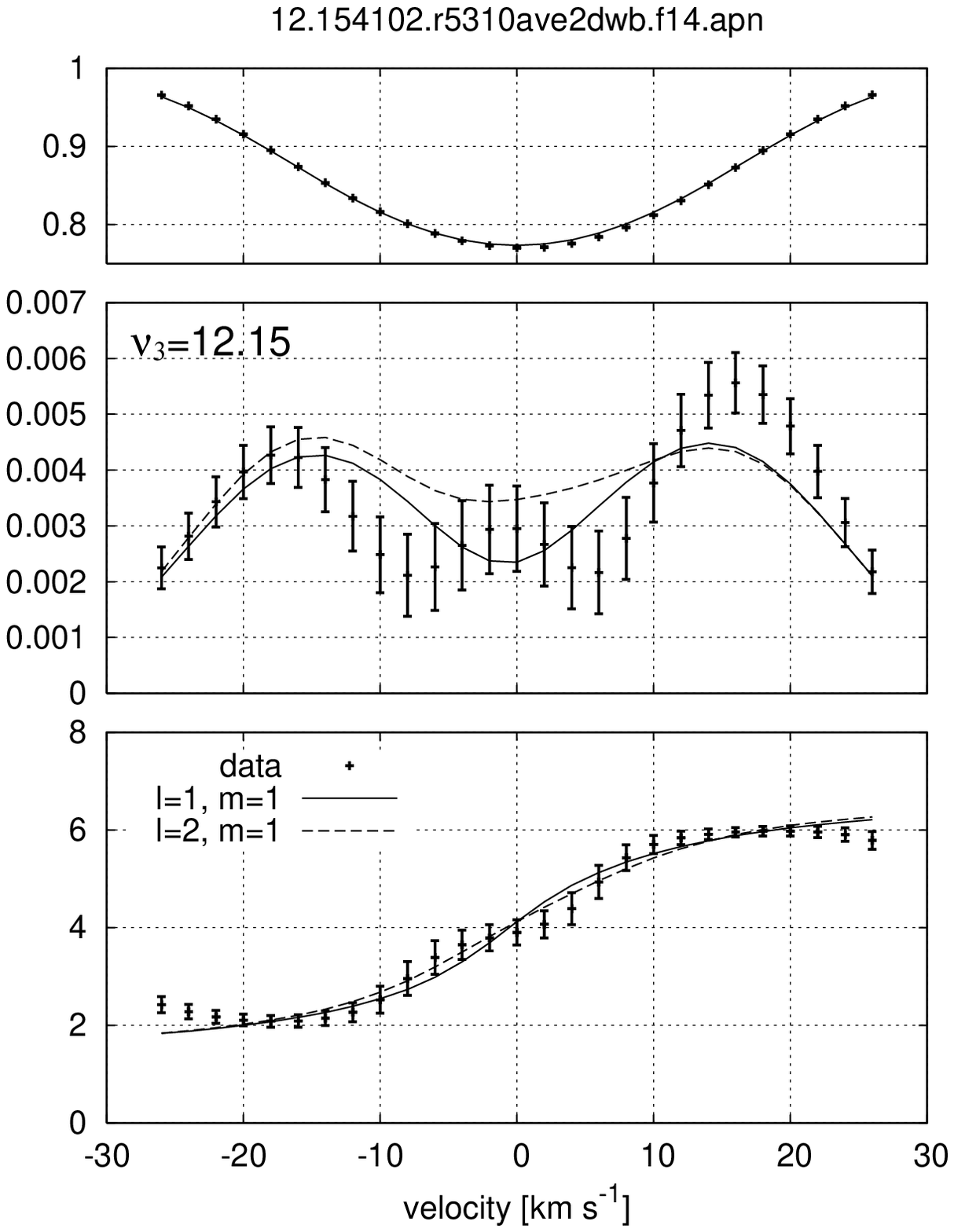}\\
\vspace*{-2mm}
  \includegraphics*[width=50mm,bb=73 104 465 435,clip,angle=0]{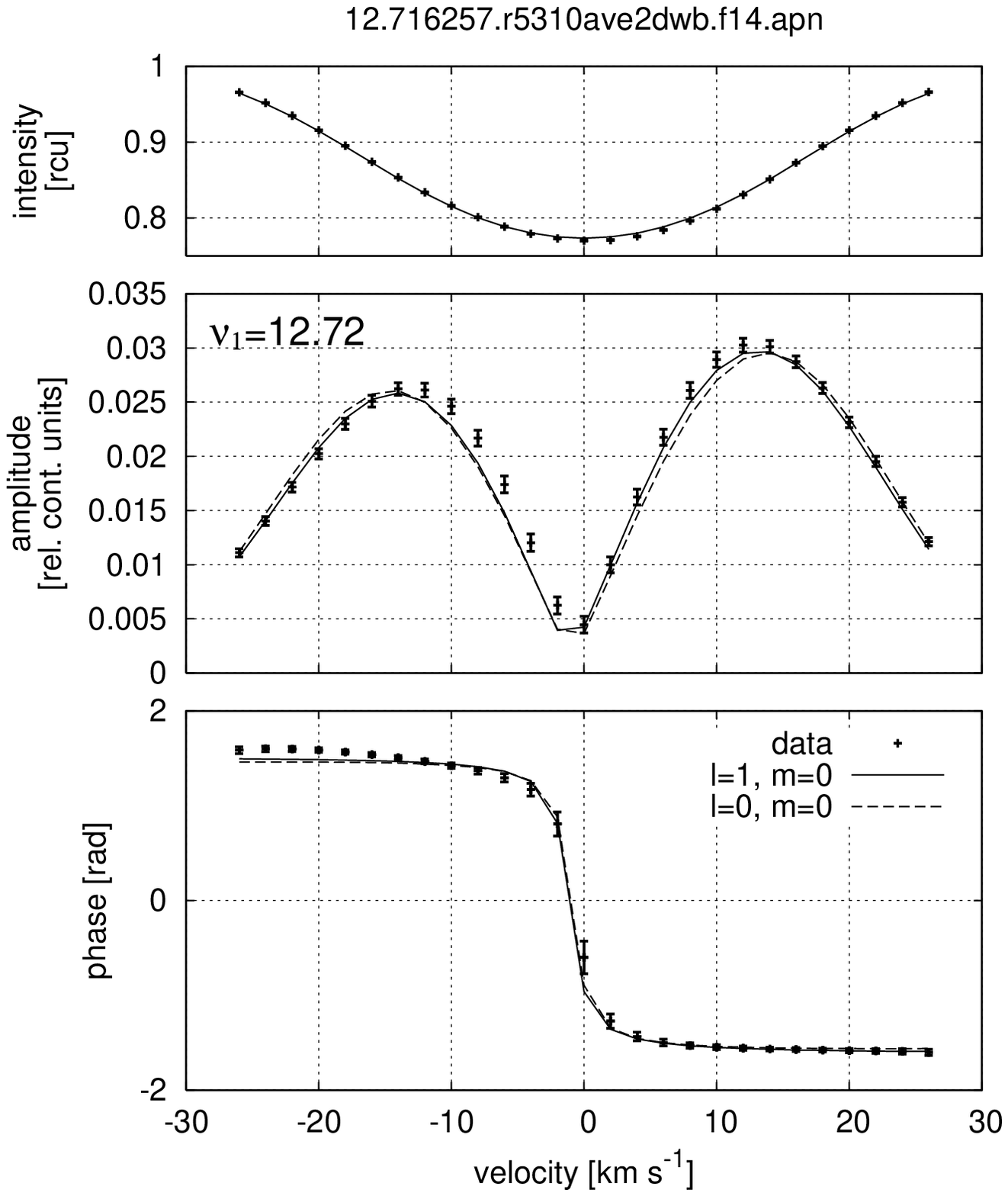}&
  \includegraphics*[width=45mm,bb=107 104 465 435,clip,angle=0]{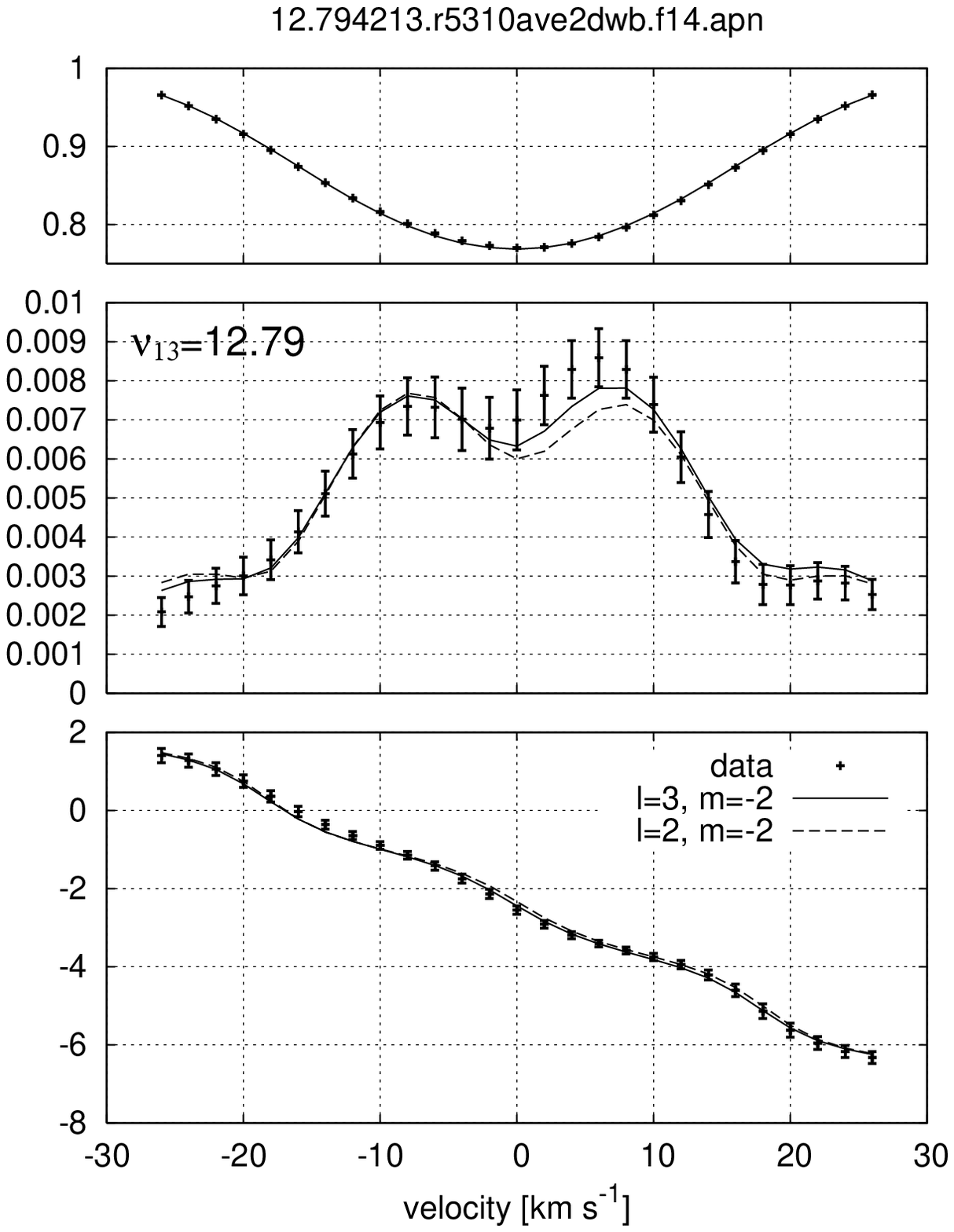} &
  \includegraphics*[width=45mm,bb=107 104 465 435,clip,angle=0]{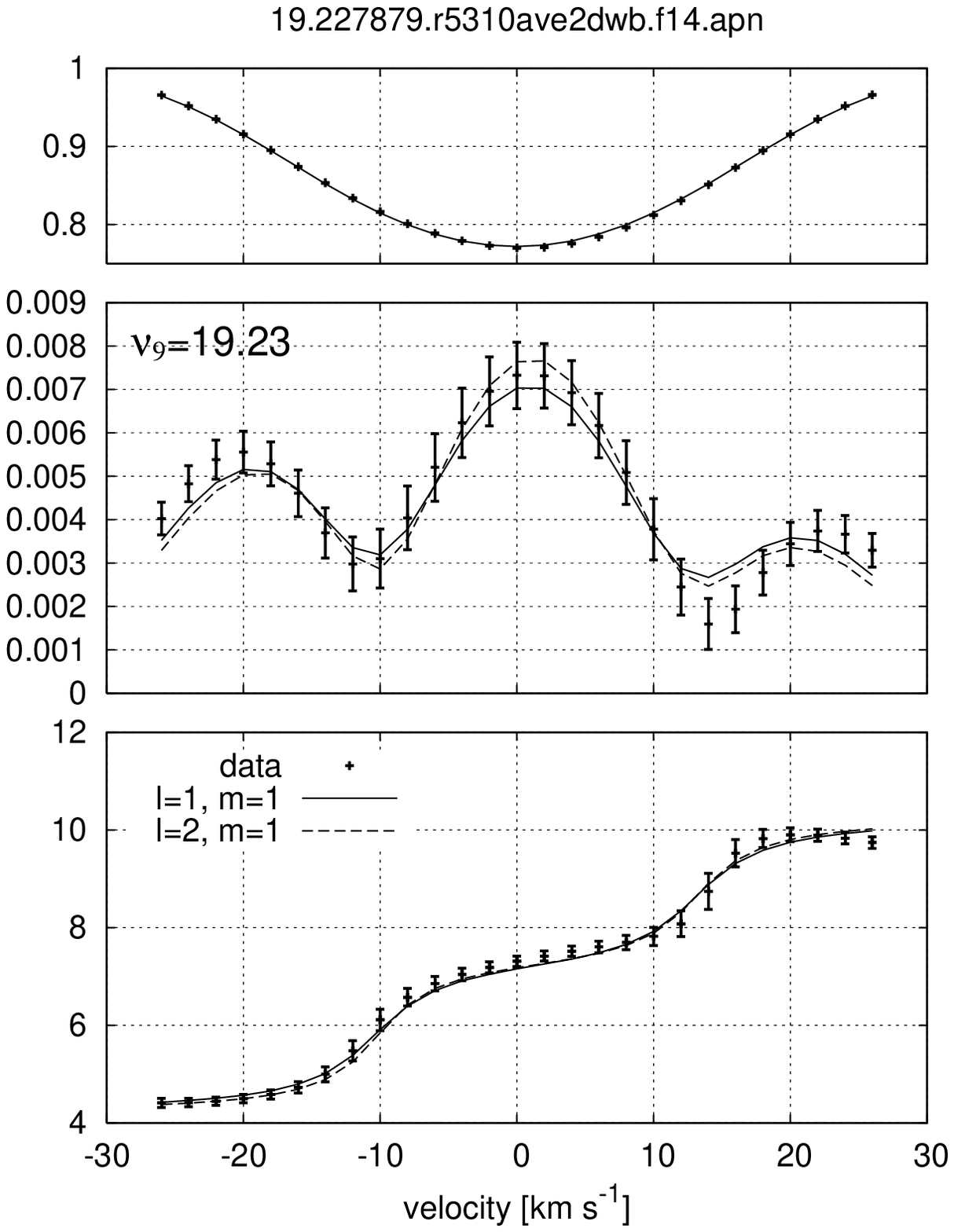}\\
\vspace*{-2mm}
  \includegraphics*[width=50mm,bb=73 104 465 435,clip,angle=0]{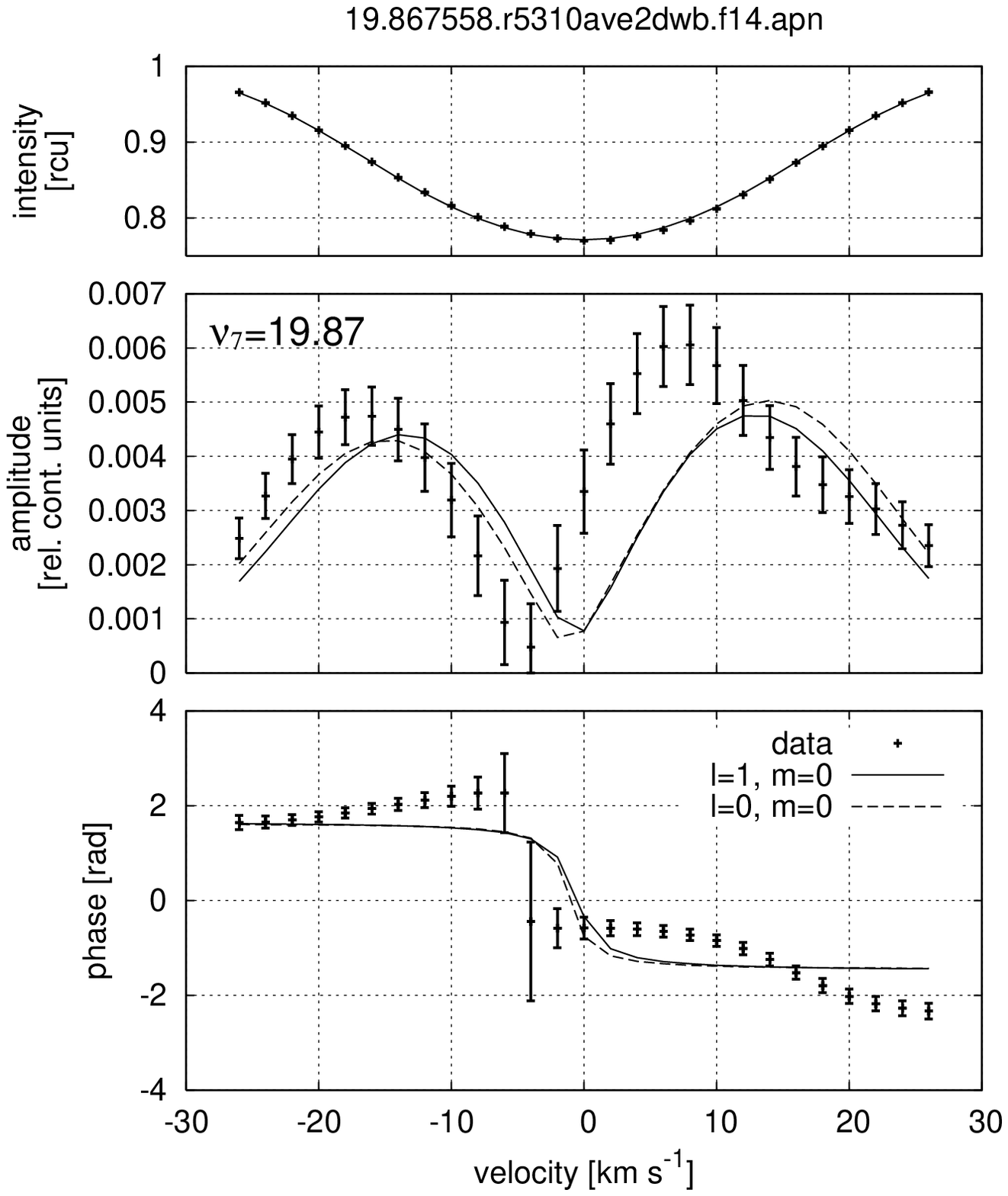} &
  \includegraphics*[width=45mm,bb=107 104 465 435,clip,angle=0]{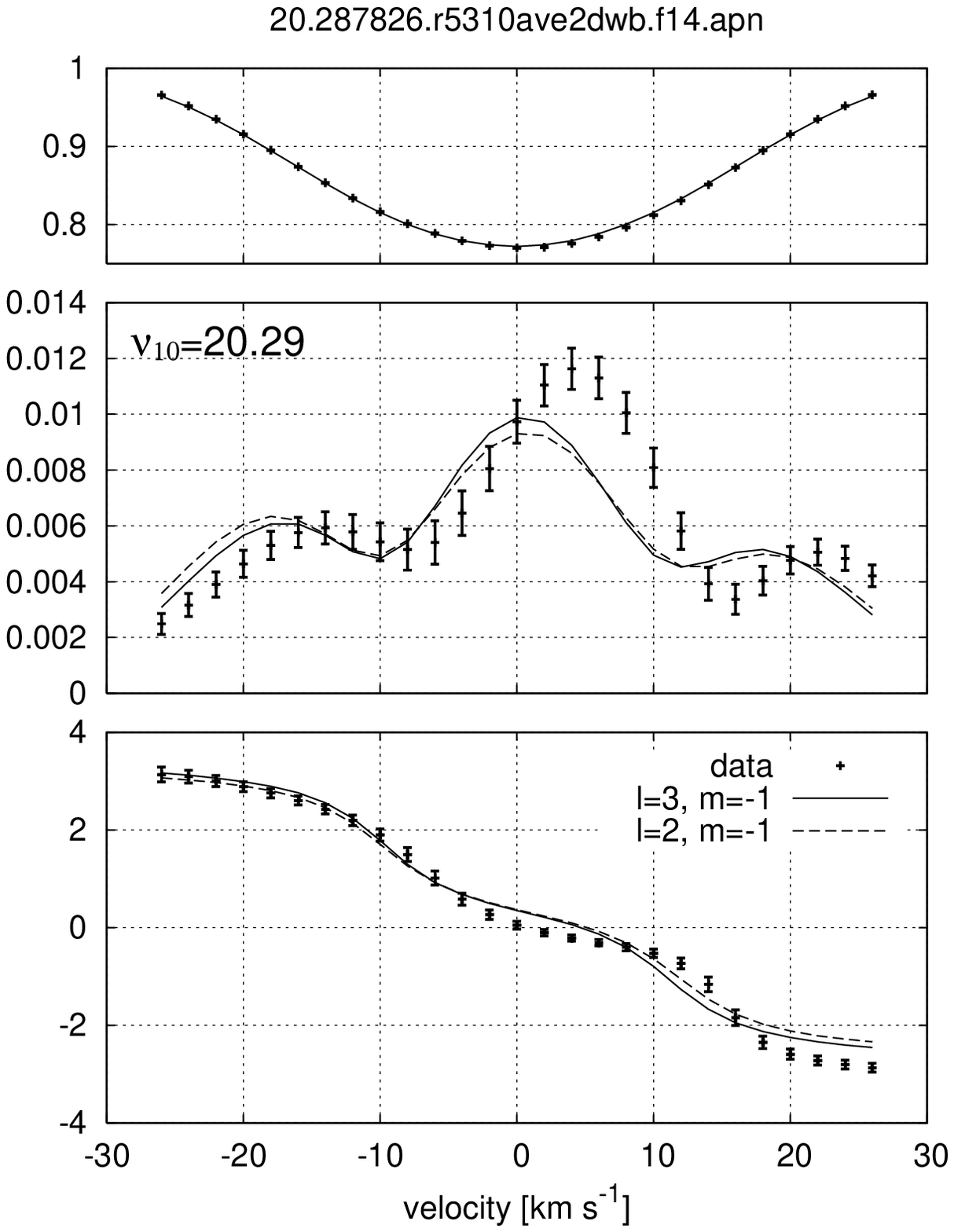}&
  \includegraphics*[width=45mm,bb=107 104 465 435,clip,angle=0]{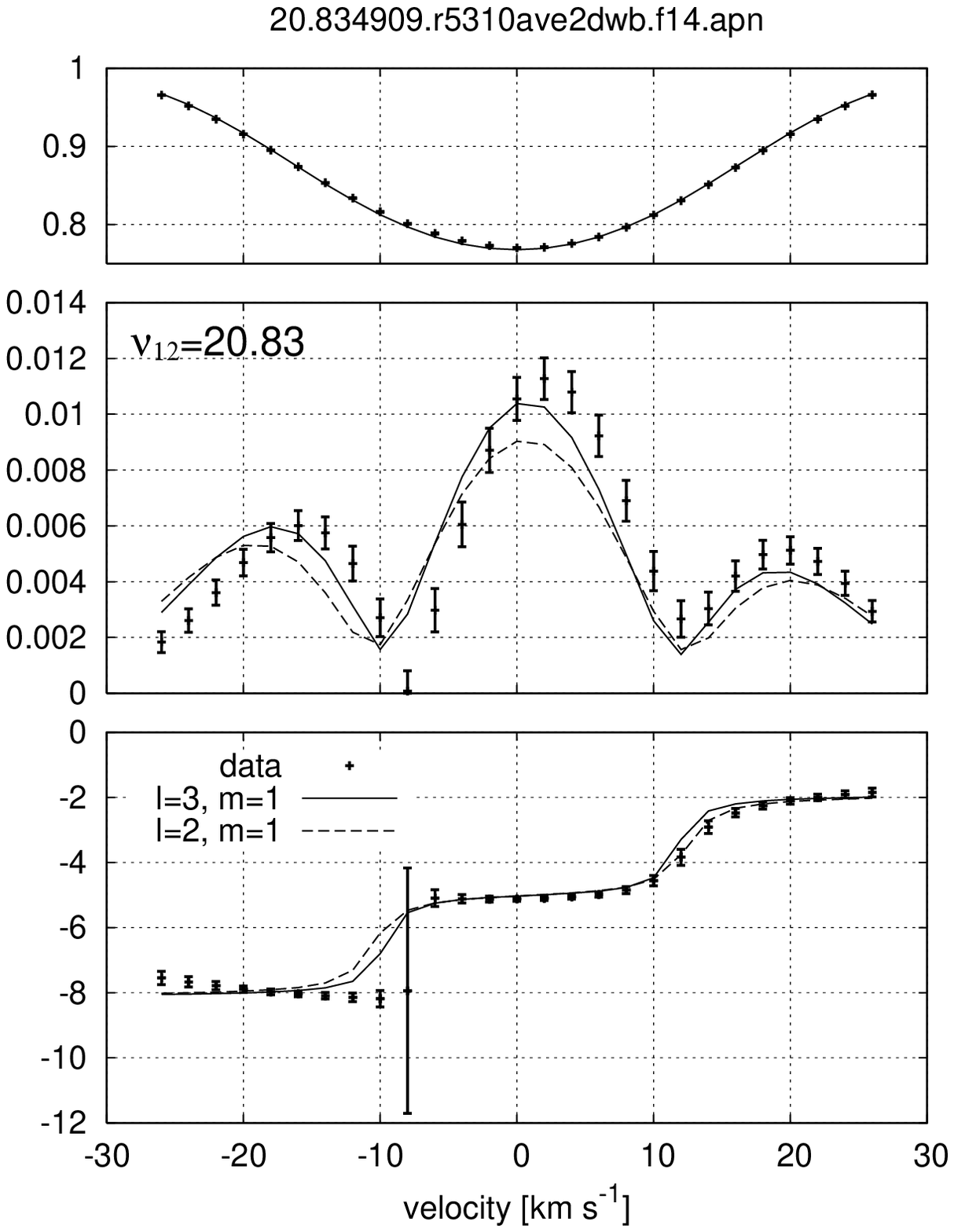}\\
\vspace*{-2mm}
  \includegraphics*[width=50mm,bb=73 104 465 435,clip,angle=0]{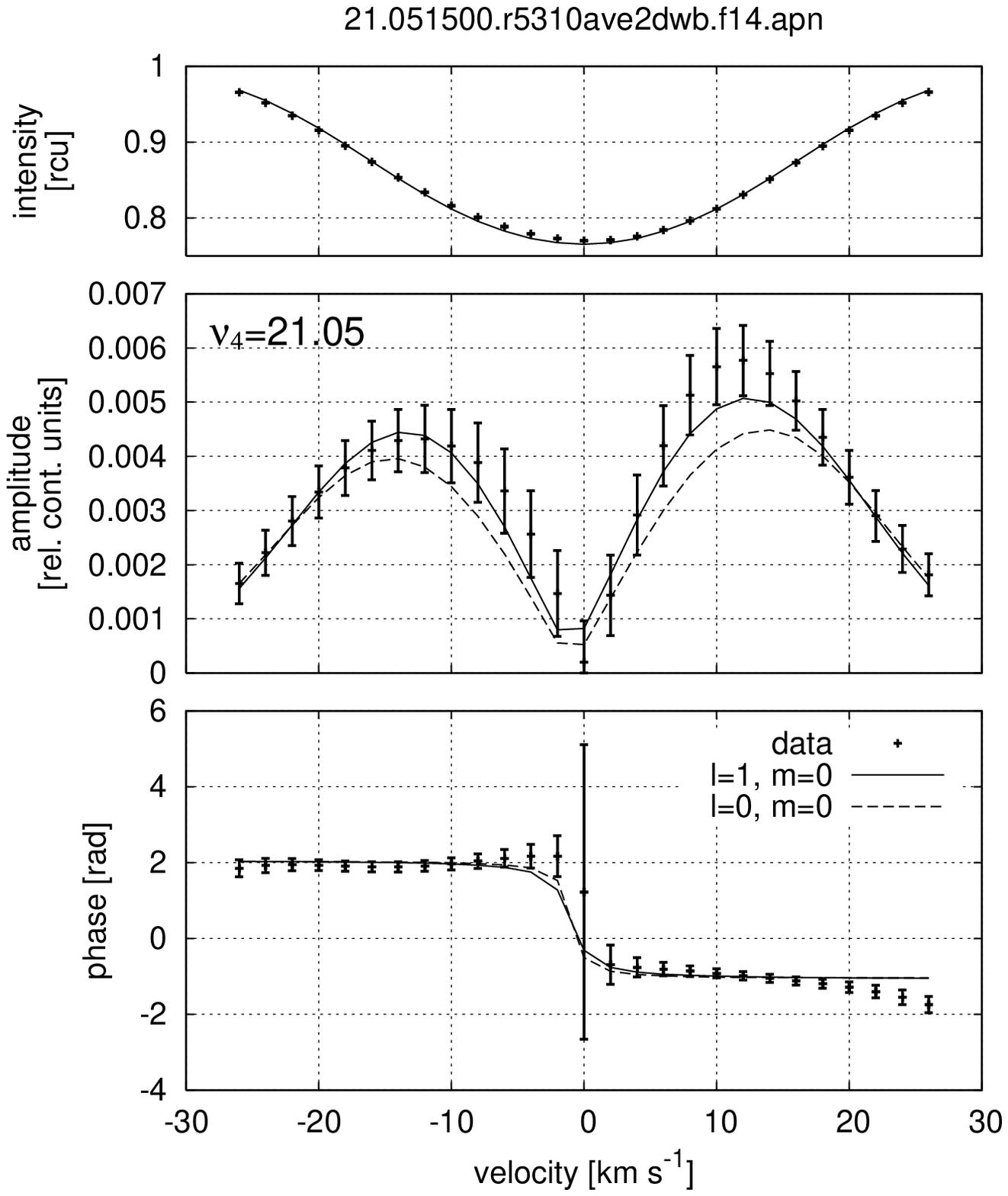} &
  \includegraphics*[width=45mm,bb=107 104 465 435,clip,angle=0]{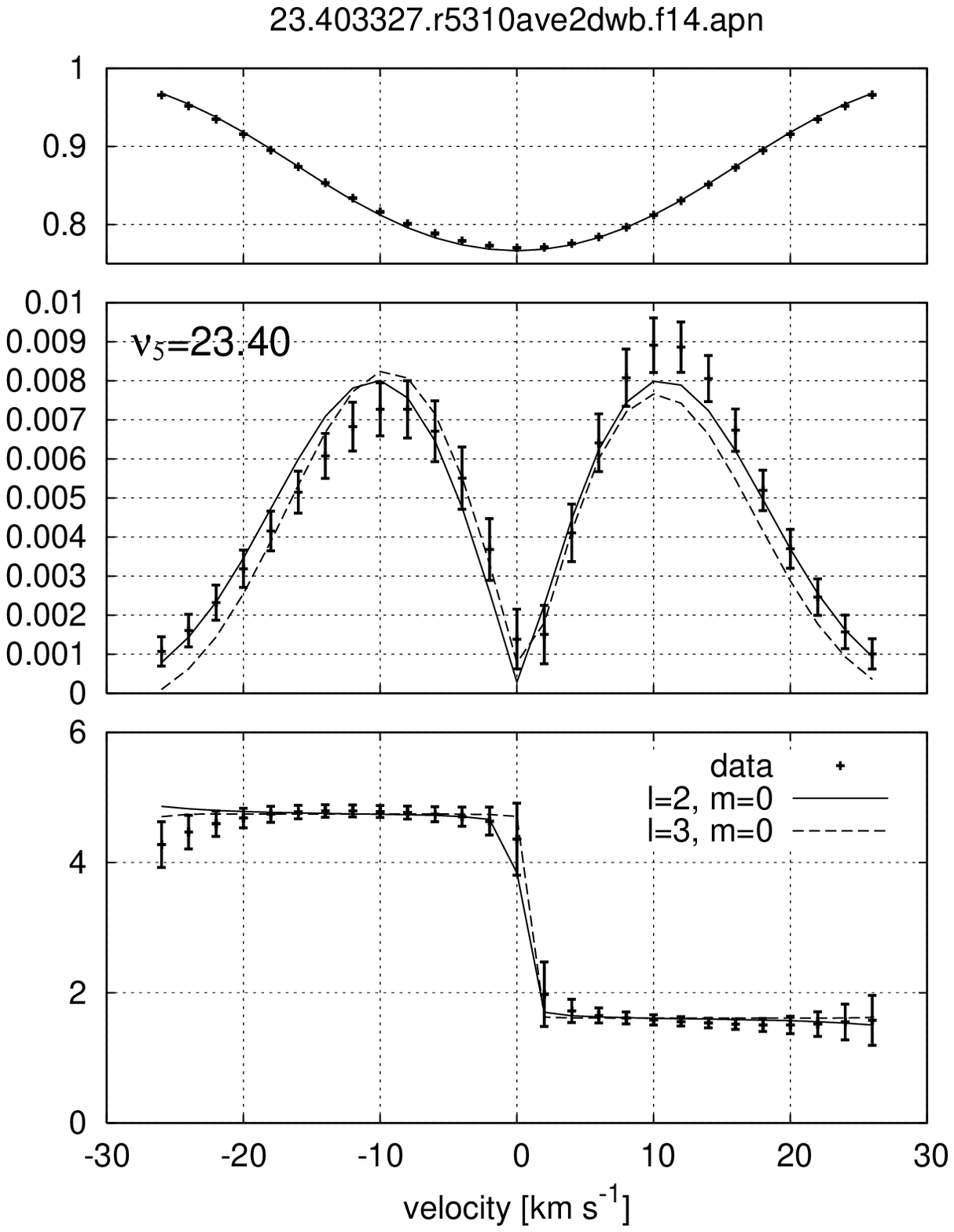} &
  \includegraphics*[width=45mm,bb=107 104 465 435,clip,angle=0]{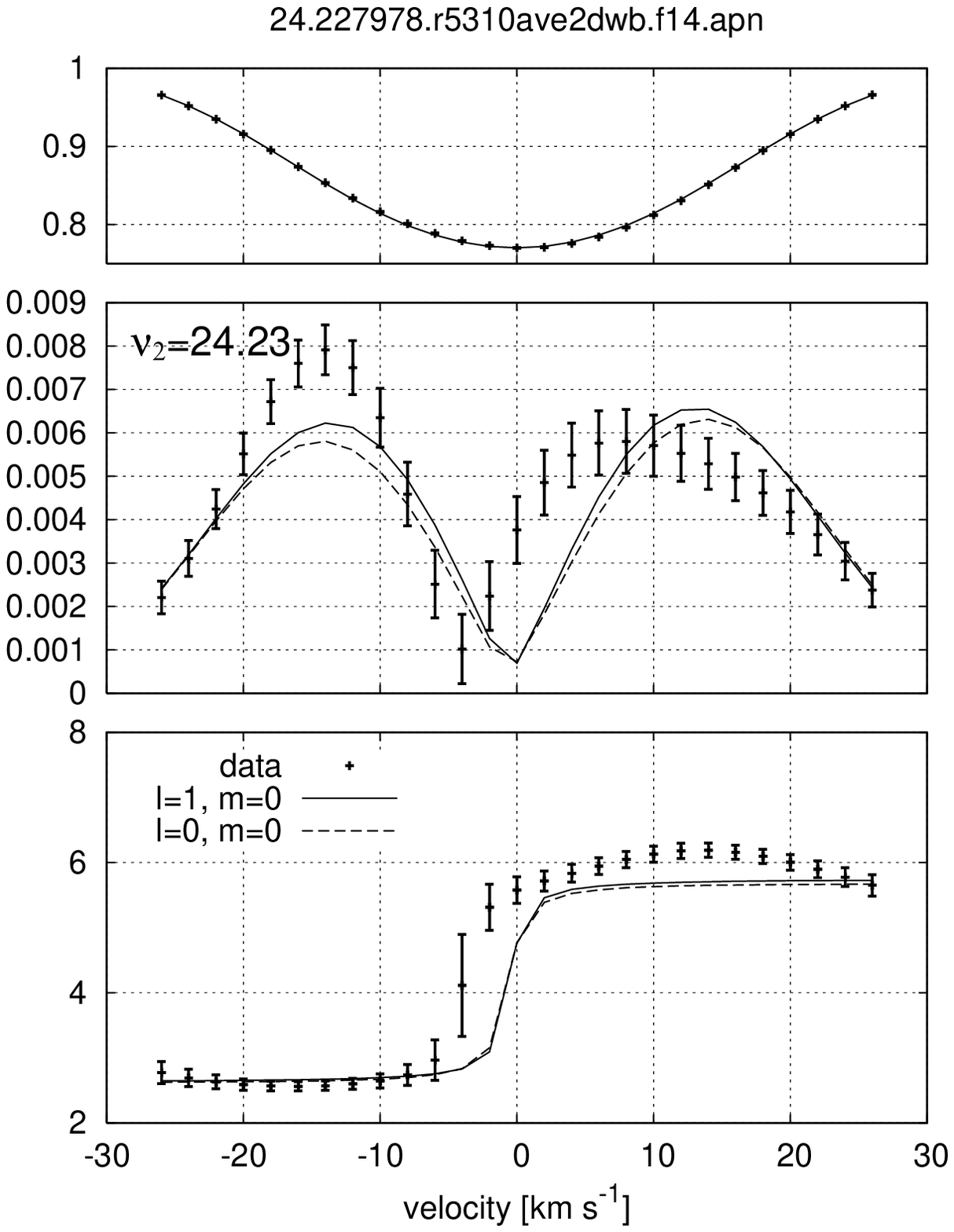}\\
\vspace*{-2mm}
  \includegraphics*[width=50mm,bb=73 80 465 435,clip,angle=0]{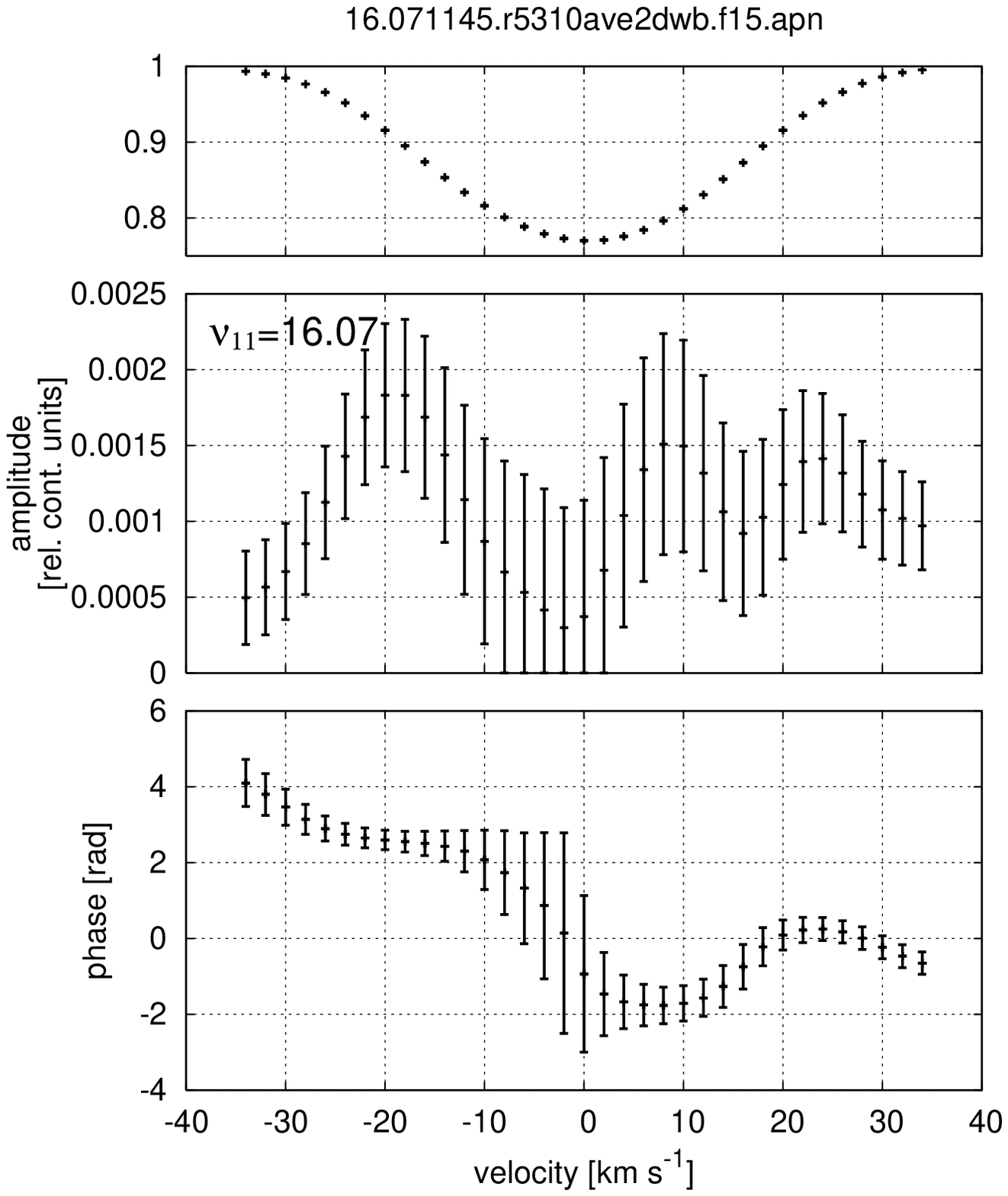} &
  \includegraphics*[width=45mm,bb=107 80 465 435,clip,angle=0]{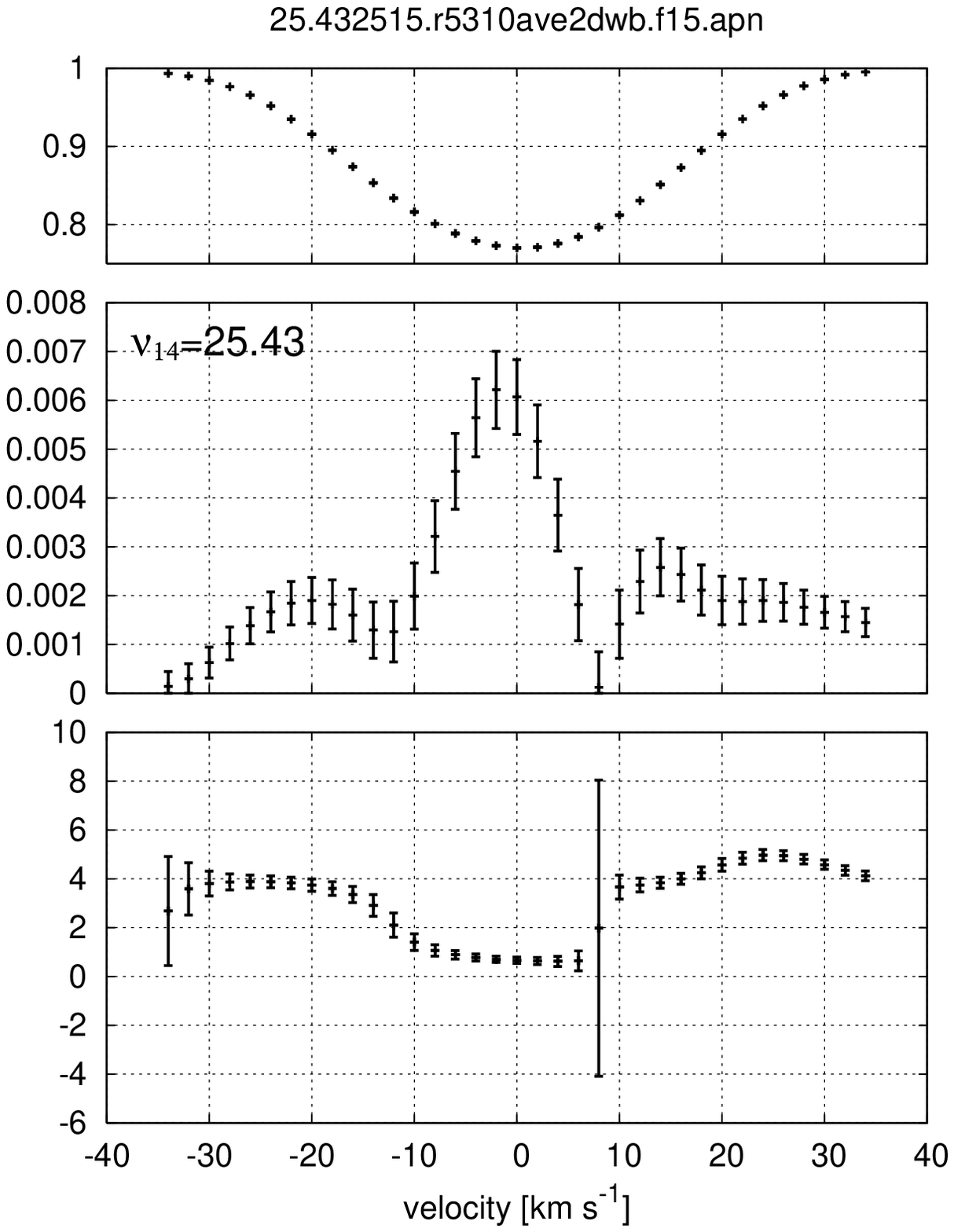} &
  \includegraphics*[width=45mm,bb=107 80 465 435,clip,angle=0]{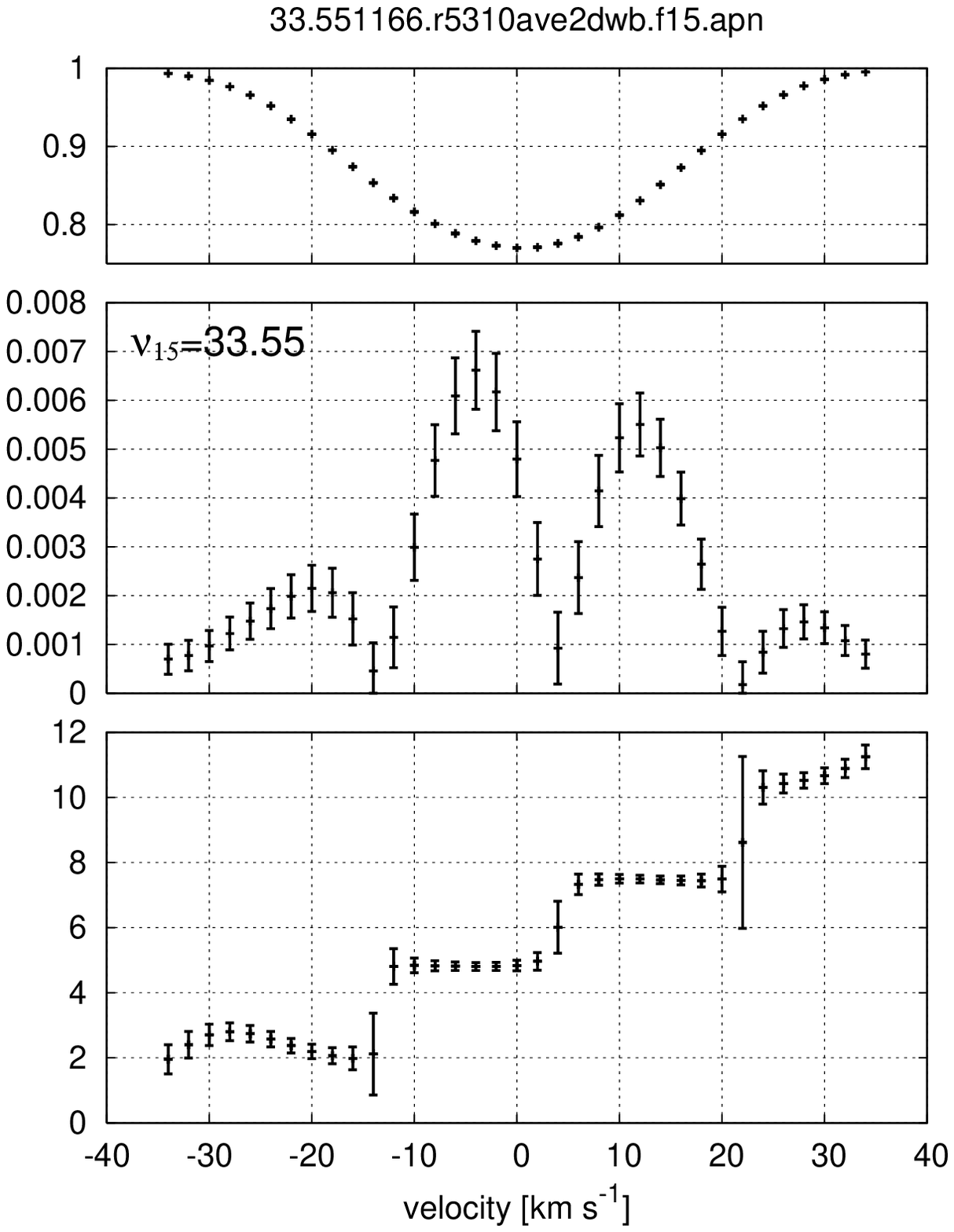}\\
\end{tabular}
\end{center}
\vspace*{-5mm}
\caption{Fit of the Fourier parameters across the line profile of the detected pulsation modes taking equivalent width variations into account. The top panels show the zero-point profile. For every single frequency, the observed amplitude in units of the continuum and the phase distribution in radians is shown together with the two best fitting models. For the three frequencies shown in the bottom panels, no fits have been computed due to the low S/N (in the case of $\nu_{11}=16.07$~\cd) and due to their nature as harmonic ($\nu_{14}=25.43$~\cd) or combination frequency ($\nu_{15}=33.55$~\cd), respectively.}
\label{fig:fitzap2}
\end{figure*}
\subsection{Mode identification by means of the FPF method}
For pinpointing the properties of the pulsation modes we applied the FPF method. By means of this method, the observed amplitude and phase values across the profile are fitted with theoretical values derived from synthetic line profiles. The fitting is carried out by applying genetic optimization routines in a large parameter space. The reduced $\chi^2_\nu$ is calculated from complex amplitudes in order to combine amplitude and phase information in the following way 
\begin{equation}
\chi^2_\nu=\frac{1}{2n_\lambda-m}\sum^{n_\lambda}_{i=1}\Biggl[\frac{(A^o_{R,i}-A^t_{R,i})^2}{\sigma^2_{R,i}}+\frac{(A^o_{I,i}-A^t_{I,i})^2}{\sigma^2_{I,i}}\Biggr]
\label{eq:chisq}
\end{equation}
where $n_\lambda$ is the number of pixels across the profile, $m$ is the number of free parameters, $A^o$ and $A^t$ denote observationally and theoretically determined values, respectively, $A_R=A_\lambda \cos\phi_\lambda$ and $A_I=A_\lambda \sin\phi_\lambda$ are the real and imaginary part of the complex amplitude, and $\sigma^2$ is the observed variance. Eq. (\ref{eq:chisq}) can be easily modified if the observed amplitude and phase from photometric passbands should be included for the calculation of $\chi^2_\nu$.

Since the amplitude and phase of a given wavelength bin are treated as independent variables, the variances are calculated from
\begin{equation}\sigma^2_{R,\lambda}=\sigma(A_\lambda)^2 \cos^2 \phi_\lambda +\sigma(\phi_\lambda)^2 A^2_\lambda \sin^2 \phi_\lambda\end{equation}
\begin{equation}\sigma^2_{I,\lambda}=\sigma(A_\lambda)^2 \sin^2 \phi_\lambda +\sigma(\phi_\lambda)^2 A^2_\lambda \cos^2 \phi_\lambda\end{equation}

We describe the theoretical stellar surface displacement field as superposition of spherical harmonics assuming linear pulsation theory. The description is limited to a slowly rotating system taking into account toroidal motion caused by the Coriolis force. Such a description has also been applied by Aerts \& Waelkens (1993) in the case of the MM and by Schrijvers et al. (1997). The intrinsic unbroadened line profile was assumed to be a Gaussian. The broadened profile was computed by summing over a 7200-element grid approximating the visible distorted stellar surface. A quadratic limb darkening law was adopted. For more details on the computation of the theoretical LPV refer to Appendix A of Paper I.

The observed uncertainties $\sigma(A_\lambda)$ and $\sigma(\phi_\lambda)$ were computed from the standard deviation of the pixels across the line after prewhitening of all significant frequencies. The behavior of the observed zero point, amplitude, and phase of all 15 detected terms is shown in Figure~\ref{fig:fitzap2}. The observed values are denoted by the crosses with error bars, whereas the lines denote fits described below. The top-most panel is the zero point profile, which is identical for all frequencies. For each frequency's diagram, the upper panel shows the amplitude in units relative to the continuum and the panel below shows the phase distribution expressed in radians.

A visual inspection on these diagrams already reveals a lot of interesting information about the mode properties.  
\begin{itemize}
\item The dominant mode $\nu_1=12.72$~\cd~shows clearly an axisymmetric behavior ($m=0$), namely a double-peaked amplitude distribution with zero amplitude at the center of the profile accompanied by a phase shift of $\pi$ at the same position. The difference of the height of the two amplitude maxima is intrinsic to the LPV. It is caused by the variable equivalent width due to the temperature variations. 
\item The double-peaked amplitude distribution of five other modes, namely $\nu_{8}=9.66$~\cd, $\nu_{7}=19.87$~\cd, $\nu_{4}=21.05$~\cd, $\nu_{5}=23.40$~\cd, and $\nu_{2}=24.23$~\cd, also shows the characteristics of axisymmetric modes. Here, differences in the heights of the two maxima might be due to observed uncertainties considering the error bars. The phase-jump of $\pi$ at the minimum amplitude is also consistent with modes of $m=0$.
\item This high number of axisymmetric modes is in agreement with the observations reported by MP02, who have identified 8 out of 10 detected modes as $m=0$. The reason for this might be a relatively low inclination angle combined with the low \vsini~of FG Vir favoring the detection of axisymmetric modes, which are best seen pole-on. Since the rotationally broadened line profile represents an one-dimensional projection of the stellar surface, higher values of \vsini result in spatially better resolution. Therefore, high-degree modes are easier detectable at high \vsini and FG Vir at about \vsini=21~\kms is not expected to show spectroscopically high-degree modes, just as it is observed.
\item At least five modes are non-axisymmetric: $\nu_6=9.20$~\cd, $\nu_9=19.23$~\cd, $\nu_{10}=20.29$~\cd, $\nu_{12}=20.83$~\cd, and $\nu_{13}=12.79$~\cd. This is deduced from their amplitude distribution, which has a maximum close to the line center, combined with a gradual phase change across the line of more than $\pi$ and therefore an azimuthal order $|m|>0$. $\nu_6$, $\nu_9$ and $\nu_{12}$ appear to be retrograde modes ($m>0$) due to the gradual increase in phase from blue to red, whereas $\nu_{13}$ and $\nu_{10}$ are apparently prograde modes.
\item The diagrams of $\nu_3=12.15$~\cd~do not permit a conclusion about its character. Photometric mode identification (Breger et al. 1999) and the results by Daszynska-Daszkiewicz et al (2005), hereafter refered to as DD05, strongly point towards the identification as fundamental radial pulsation mode. From our derived amplitudes and phases across the line, we cannot confirm these results. A reason might be that in our spectroscopic data it is impossible to separate $\nu_3$ from a photometrically detected close frequency at 12.162~\cd. An interaction of these two modes in Fourier space could prevent a successful mode identification.
\item Due to the low S/N of $\nu_{11}=16.07$~\cd~we exclude this mode from our mode identification. This includes also the harmonic and combination terms, $\nu_{14}$ and $\nu_{15}$ respectively, since these are not intrinsic pulsation modes.
\end{itemize}

\begin{figure}[!ht]
\centering
  \includegraphics*[height=80mm,bb=78 28 570 720,clip,angle=-90]{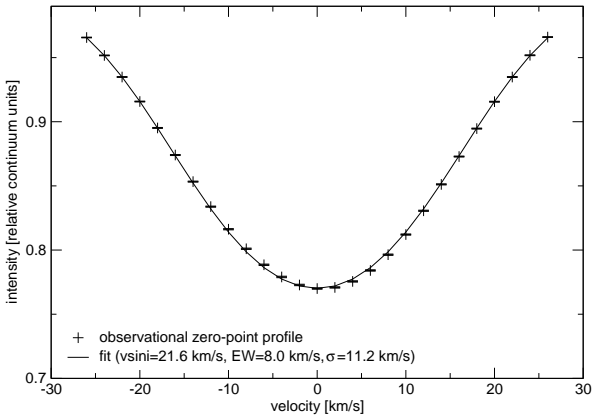}
  \caption{Observed zero point profile and best fit by means of a synthetic rotationally broadened profile.}
 \label{fig:vsinifit}
\end{figure}

Our modeled line profiles depend on the stellar parameters: mass $M$, \teff, \logg, limb darkening, and on the flux derivatives, $\alpha_T$ and $\alpha_g$. We adopted the mean stellar parameters derived by DD05. Photometric calibrations of Kurucz models as well as Vienna atmospheres yield consistent values of $\log$~\teff = 3.87$\pm$0.01 and \logg=4.0$\pm$0.1. DD05 determined an improved value of \teff~by fitting the amplitude and phase of the dominant mode $\nu_1$ to the observations using flux derivatives determined from NEMO.2003 models (Nendwich et al. 2004). The best fit was achieved for a model of solar metal abundance and a micro-turbulence velocity of $\xi=4$~\kms, yielding a value of $\log$~\teff~=3.86 and \logg=4.0. Our flux derivatives $\alpha_T$ and $\alpha_g$ were computed from NEMO.2003 models choosing the previously determined values of \teff~and \logg. 

Our limb-darkening coefficients were taken from tabular data provided by Barban et al. (2003). We adopted the values for the quadratic limb-darkening law for the Str\"omgren filters $v$ and $y$.
We excluded the wings of the line from our analysis due to the fact that there the rotationally broadened profile is dominated by the Lorentz profile deviating from the adopted Gaussian intrinsic profile. Therefore, our fit covers only the range between -26 and 26~\kms. 

The projected rotational velocity \vsini, the width of the intrinsic profile $\sigma$, and the equivalent width $W$ were computed from a least-squares fit of a rotationally broadened synthetic profile to the zero point profile (see Fig.~\ref{fig:vsinifit}). We obtained the following values: \vsini=21.6$\pm$0.3~\kms, $W$=8.02$\pm$0.05~\kms, $\sigma=11.2\pm0.3$~\kms. The zero point profile is slightly broadened due to the LPV, mainly by the contribution of $\nu_1$, which has RV/\vsini=0.1. The resulting overestimation of \vsini~is only about 1\%.

\begin{table*}[!ht]
\centering
\normalsize
\caption{Mode parameters derived from the application of the FPF method by including equivalent width variations. For each pulsation mode, the five best solutions are shown. $a_s$ is the relative displacement amplitude; $i$ is the stellar inclination angle in degrees; \vsini~is the projected rotational velocity, $\sigma$ is the width of the intrinsic profile, both expressed in \kms; $|f|$ and $\psi_f$ are the ratio and the phase shift, respectively, between flux and radius variations. \alphaw~describes the dependence of the equivalent width with respect to the temperature. The 95~\% significance limit is at $\chi^2_\nu=1.37$. Due to spherical symmetry we have not indicated the inclination angle for radial modes.}
\scriptsize
\begin{tabular}{|@{}c@{}|c|c|c|c|c|c|c|c|c||@{}c@{}|c|c|c|c|c|c|c|c|c|}
\hline
\multicolumn{10}{|c||}{$\nu_6=9.199$~\cd} & \multicolumn{10}{|c|}{$\nu_8=9.656$~\cd}\\ \hline
$\chi^2_\nu$ & $\ell$ & $m$ & $a_s$ &$i$ & $v\sin i$ & $\sigma$ & $|f|$ & $\psi_f$ & \alphaw & $\chi^2_\nu$ & $\ell$ & $m$ & $a_s$ &$i$ & $v\sin i$ & $\sigma$ & $|f|$ & $\psi_f$ & \alphaw\\ \hline
0.66 & 3 & 1 & 0.00077 & 25 & 21.4 & 11.3 & 11.2 & 0.83 & 1.7 & 2.14 & 2 & 0  & 0.00063 & 41 & 21.9 & 11.5 & 18.0 & 0.76 & 1.7\\
0.75 & 2 & 1 & 0.00076 & 17 & 21.3 & 10.9 & 6.3  & 1.76 & 1.0 & 2.99 & 3 & 0  & 0.00033 & 10 & 21.9 & 11.5 & 17.7 & 0.68 & 5.0\\
0.95 & 1 & 1 & 0.00128 & 17 & 21.3 & 10.9 & 4.9  & 1.64 & 1.7 & 3.01 & 1 & 0  & 0.00048 & 56 & 21.3 & 10.9 & 17.1 & 0.71 & 0.3\\
4.95 & 3 & 2 & 0.00131 & 75 & 21.5 & 11.1 & 13.4 & 0.93 & 2.3 & 3.33 & 0 & 0  & 0.00042 & -  & 21.3 & 10.9 & 11.1 & 2.67 & -1.7\\
5.76 & 2 & 2 & 0.00094 & 45 & 21.8 & 11.3 & 13.7 & 1.61 & -1.7 & 4.97 & 3 & -1 & 0.00127 & 85 & 21.8 & 11.5 & 11.1 & 5.27 & 3.7\\
\hline\hline
\multicolumn{10}{|c||}{$\nu_3=12.154$~\cd} & \multicolumn{10}{|c|}{$\nu_1=12.716$~\cd}\\ \hline
$\chi^2_\nu$ & $\ell$ & $m$ & $a_s$ &$i$ & $v\sin i$ & $\sigma$ & $|f|$ & $\psi_f$ & \alphaw & $\chi^2_\nu$ & $\ell$ & $m$ & $a_s$ &$i$ & $v\sin i$ & $\sigma$ & $|f|$ & $\psi_f$ & \alphaw\\ \hline
1.85 & 1 & 1 & 0.00033 & 85 & 21.9 & 11.5 & 14.3 & 2.83 & -2.3 & 2.75 & 1 & 0 & 0.00170 & 34 & 21.3 & 11.4 & 12.0 & 2.21 & -3.0\\
2.39 & 2 & 1 & 0.00160 & 85 & 21.4 & 11.4 & 1.1  & 3.86 & 1.0 & 5.00 & 0 & 0 & 0.00212 & - & 21.3 & 10.9 & 13.1 & 2.26 & -2.3\\
4.03 & 0 & 0 & 0.00038 & -  & 21.9 & 11.5 & 7.1  & 4.54 & 2.3 & 22.62 & 2 & 0 & 0.00192 & 25 & 21.8 & 11.2 & 2.0 & 3.08 & -1.0\\
4.60 & 1 & 0 & 0.00052 & 65 & 21.5 & 11.5 & 14.3 & 3.76 & 1.7 & 97.96 & 3 & 0 & 0.00192 & 10 & 21.8 & 11.1 & 10.6 & 2.20 & -4.3\\
6.10 & 2 & 0 & 0.00028 & 25 & 21.8 & 11.5 & 15.4 & 3.12 & -1.0 & 171.81 & 1 & 1 & 0.00178 & 85 & 21.3 & 11.4 & 14.6 & 2.10 & -3.7\\
\hline\hline
\multicolumn{10}{|c||}{$\nu_{13}=12.794$~\cd} & \multicolumn{10}{|c|}{$\nu_9=19.227$~\cd}\\ \hline
$\chi^2_\nu$ & $\ell$ & $m$ & $a_s$ &$i$ & $v\sin i$ & $\sigma$ & $|f|$ & $\psi_f$ & \alphaw & $\chi^2_\nu$ & $\ell$ & $m$ & $a_s$ &$i$ & $v\sin i$ & $\sigma$ & $|f|$ & $\psi_f$ & \alphaw\\ \hline
0.56 & 3 & -2 & 0.00091 & 17 & 21.4 & 11.07 & 21.7 & 1.79 & -5.0 & 0.61 & 1 & 1 & 0.00054 & 15 & 21.9 & 11.2 & 17.8 & 1.59 & 3.0\\
0.58 & 4 & -2 & 0.00082 & 22 & 21.9 & 11.50 & 22.0 & 1.49 & -5.0 & 0.86 & 2 & 1 & 0.00036 & 17 & 21.9 & 11.5 & 20.6 & 1.39 & 3.0\\
1.01 & 2 & -2 & 0.00263 & 10 & 21.9 & 11.1 & 17.4 & 4.30 & -4.3 & 2.15 & 3 & 1 & 0.00034 & 20 & 21.9 & 11.5 & 21.0 & 1.59 & 5.0\\
1.29 & 3 & -3 & 0.00117 & 35 & 21.7 & 11.4 & 15.1 & 2.78 & -3.7 & 6.95 & 2 & 2 & 0.00033 & 60 & 21.3 & 11.4 & 16.9 & 0.59 & 4.3\\
3.76 & 3 & -1 & 0.00286 & 65 & 21.5 & 10.9 & 9.4 & 0.44 & 5.0 & 7.11 & 3 & 2 & 0.00038 & 60 & 21.6 & 11.4 & 17.4 & 1.46 & 4.3\\
\hline\hline
\multicolumn{10}{|c||}{$\nu_7=19.867$~\cd} & \multicolumn{10}{|c|}{$\nu_{10}=20.287$~\cd}\\ \hline
$\chi^2_\nu$ & $\ell$ & $m$ & $a_s$ &$i$ & $v\sin i$ & $\sigma$ & $|f|$ & $\psi_f$ & \alphaw & $\chi^2_\nu$ & $\ell$ & $m$ & $a_s$ &$i$ & $v\sin i$ & $\sigma$ & $|f|$ & $\psi_f$ & \alphaw\\ \hline
7.10  & 0 & 0  & 0.00022 & -  & 21.3 & 11.4 & 18.5 & 2.29 & -4.3 & 5.87 & 3 & -1 & 0.00045 & 29 & 21.9 & 11.5 & 22.0 & 1.31 & 4.3\\
7.12  & 1 & 0  & 0.00026 & 58 & 21.6 & 11.5 & 17.8 & 2.42 & -5.0 & 6.43 & 2 & -1 & 0.00039 & 17 & 21.3 & 10.9 & 22.0 & 1.03 & 2.3\\
7.85  & 2 & 0  & 0.00019 & 22 & 21.8 & 11.4 & 15.4 & 0.03 & 5.0 & 7.02 & 1 & -1 & 0.00065 & 17 & 21.3 & 10.9 & 22.0 & 1.69 & -0.3\\
9.21  & 3 & -1 & 0.00089 & 85 & 21.7 & 11.5 & 10.6 & 0.34 & -2.3 & 9.21 & 3 & -3 & 0.00033 & 75 & 21.7 & 11.5 & 11.1 & 0.49 & 1.0\\
10.10 & 2 & -2 & 0.00014 & 80 & 21.9 & 11.4 & 11.1 & 5.81 & 2.3 & 33.21 & 3 & 1 & 0.00024 & 10 & 21.9 & 10.9 & 17.7 & 1.17 & 5.0\\
\hline\hline
\multicolumn{10}{|c||}{$\nu_{12}=20.834$~\cd} & \multicolumn{10}{|c|}{$\nu_4=21.051$~\cd}\\ \hline
$\chi^2_\nu$ & $\ell$ & $m$ & $a_s$ &$i$ & $v\sin i$ & $\sigma$ & $|f|$ & $\psi_f$ & \alphaw & $\chi^2_\nu$ & $\ell$ & $m$ & $a_s$ &$i$ & $v\sin i$ & $\sigma$ & $|f|$ & $\psi_f$ & \alphaw\\ \hline
4.14  & 3 & 1  & 0.00033 & 10 & 21.5 & 10.9 & 17.7 & 1.32 & 5.0 & 0.81 & 1 & 0 & 0.00039 & 70 & 21.3 & 10.9 & 15.0 & 2.32 & -3.7\\
4.61  & 4 & 1  & 0.00040 & 10 & 21.9 & 11.5 & 22.0 & 1.59 & 5.0 & 1.06 & 0 & 0 & 0.00022 & -  & 21.3 & 10.9 & 11.2 & 2.32 & -4.3\\
5.99  & 2 & 1  & 0.00033 & 10 & 21.3 & 10.9 & 17.1 & 1.27 & 2.3 & 1.20 & 2 & 0 & 0.00025 & 34 & 21.9 & 11.5 & 15.0 & 2.32 & -5.0\\
8.34  & 1 & 1  & 0.00054 & 10 & 21.3 & 10.9 & 18.9 & 0.96 & 3.7 & 3.42 & 3 & 0 & 0.00019 & 10 & 21.9 & 11.5 & 22.0 & 2.39 & -5.0\\
11.62 & 3 & -1 & 0.00033 & 10 & 21.7 & 10.9 & 14.3 & 1.71 & 1.0 & 6.57 & 1 & 1 & 0.00019 & 85 & 21.3 & 11.3 & 15.7 & 2.05 & -1.7\\
\hline\hline
\multicolumn{10}{|c||}{$\nu_5=23.403$~\cd} & \multicolumn{10}{|c|}{$\nu_2=24.227$~\cd}\\ \hline
$\chi^2_\nu$ & $\ell$ & $m$ & $a_s$ &$i$ & $v\sin i$ & $\sigma$ & $|f|$ & $\psi_f$ & \alphaw & $\chi^2_\nu$ & $\ell$ & $m$ & $a_s$ &$i$ & $v\sin i$ & $\sigma$ & $|f|$ & $\psi_f$ & \alphaw\\ \hline
0.74  & 2 & 0 & 0.00040 & 41 & 21.3 & 11.0 & 22.0 & 1.46 & -1.0 & 3.72 & 1 & 0 & 0.00093 & 83 & 21.6 & 10.9 & 10.1 & 2.67 & -5.0\\
1.22  & 3 & 0 & 0.00025 & 10 & 21.9 & 11.5 & 21.7 & 1.61 & -1.0 & 3.94 & 0 & 0 & 0.00025 & - & 21.3 & 10.9 & 15.4 & 2.60 & -3.7\\
4.24  & 1 & 0 & 0.00037 & 66 & 21.3 & 10.9 & 22.0 & 1.08 & -1.0 & 5.32 & 2 & 0 & 0.00019 & 10 & 21.3 & 11.5 & 2.1 & 0.98 & 1.7\\
5.99  & 0 & 0 & 0.00025 & -  & 21.3 & 10.9 & 10.1 & 1.56 & -2.3 & 9.83 & 1 & 1 & 0.00019 & 85 & 21.7 & 11.5 & 13.4 & 4.00 & 2.3\\
12.40 & 1 & 1 & 0.00019 & 80 & 21.3 & 11.2 & 16.6 & 2.20 & -0.3 & 10.20 & 3 & 0 & 0.00022 & 10 & 21.9 & 11.5 & 22.0 & 3.20 & -1.7\\
\hline
\end{tabular}
\label{tab:mi2_1}
\vspace*{5mm}
\end{table*}

\subsection{Results of the mode identification by neglecting equivalent width variations}
We started our mode identification by neglecting the equivalent width variations of the spectral line due to temperature variations. This was achieved by setting $\alpha_W=0$. A genetic optimization (for details see Paper I) with the following free parameters [min, max; step] was adopted: $\ell$=[0, 4; 1], $m$=[-$\ell$, $\ell$; 1], the relative displacement amplitude $a_s$=[0.00005, 0.003; 0.00005], $i$=[5, 90; 5]$^\circ$, \vsini=[21.3, 21.9; 0.1]~\kms, $\sigma$=[10.9, 11.5; 0.1]~\kms. Since we know the equivalent width of the line with much better precision than \vsini~and $\sigma$, we fixed it during the optimization to 8.02~\kms. The theoretical equatorial break-up velocity of FG Vir is at about $v_{eq}=210$~\kms~(Collins \& Harrington 1966), which leads to a minimum possible inclination angle of about $6^\circ$. It therefore is reasonable to set the lower inclination limit to this value. For each pulsation mode we calculated approximately 40000 models during the process of optimization.

According to this identification, six modes are axisymmetric, four are retrograde with $m=1$ and two modes are prograde with $m=-1$ and $m=-2$ respectively. The inclination of the best fitting models covers almost the whole tested range from 10 to 85$^\circ$. 

Only four of the twelve tested modes, $\nu_4$, $\nu_5$, $\nu_6$, and $\nu_{13}$ can be fitted satisfactory in the adopted confidence limit. The $\chi^2_\nu$ of $\nu_1$ is especially high ($\chi^2_\nu=10.72$), which is mainly due to the fact that this mode clearly shows equivalent width variations which are ignored here. We will focus on this problem in detail in the next subsection, where we discuss the fits when equivalent width variations are included. Nevertheless, for nine modes the three best solutions yield identical $m$ values. The fact that $m$ is much better constrained than $\ell$ is consistent with the results of our experimental tests with synthetic data (see Paper I). 

There are some modes, especially $\nu_7=19.87$~\cd, $\nu_{10}=20.29$~\cd~and $\nu_2=24.23$~\cd, showing a significant velocity shift between the observed and modeled amplitude distribution that might be connected to equivalent width variations of these terms. Another explanation considers photometrically detected close frequency terms which distort the Fourier parameters across the line. 

\subsection{Results of the mode identification by including equivalent width and light variations}
In an improved approach we carried out a mode identification by also including the equivalent width variations in a parameterized form caused by the pulsational temperature variations. The complex parameter $f$, the ratio of the stellar flux to radius variations, governs the relative temperature change and therefore also the equivalent width change of the line. This gives us the opportunity to derive $f$ and its phase shift $\psi_f$ empirically from the observed equivalent width variations. The parameter space for the optimization was extended by varying the following values: $|f|$~[0,22;0.5], $\psi_f$~[0, 2$\pi$;0.05], the change of the equivalent width with respect to the temperature change \alphaw~[-5, 5; 1] (see Paper I for details).

We report the best fits for each mode in Tables~\ref{tab:mi2_1} and the corresponding amplitude/phase diagrams in Figure~\ref{fig:fitzap2}. The $\chi^2_\nu$ value of five modes, $\nu_4$, $\nu_5$, $\nu_6$, $\nu_9$, and $\nu_{13}$ reaches below the 95~\%-significance limit of 1.37. Now, additionally $\nu_9=19.23$~\cd~can be satisfactorily modeled as a retrograde mode of $m=1$ and $\ell=1$ or $2$. For all modes, the minimum $\chi^2_\nu$ is lower than in the first approach that did not include equivalent width variations.


Most importantly, the fit for the dominant mode $\nu_1$ is much better now and points towards a dipole mode, consistent with previous mode identifications from other authors (Breger et al. 1999, Mantegazza \& Poretti 2002). From the fit to this mode an inclination of 35$^\circ$ is derived. But, as shown in Paper I, axisymmetric modes are inappropriate for determining the inclination since almost any inclination value can be obtained by scaling the amplitude.  It is much better to rely on the non-axisymmetric modes like $\nu_6$, $\nu_9$ and $\nu_{13}$, which all point towards a low inclination value between 10 and 25$^\circ$. We will explore this in more detail in the next subsection, where an improved value of $i$ is computed by simultaneously fitting these three modes. 

Formally, we can only fit five pulsation modes satisfactorily and other modes can not be reproduced within the adopted significance limit. But also for the latter modes, some $\ell$ and $m$ must be correct. Apparently, their Fourier diagrams are distorted in such a way that we cannot model them.

In most cases, the fact that the modeled Fourier values fail to fit the observations is due to a blue or red shift of the observed amplitude and phase values. This is best seen for $\nu_2$ and $\nu_{10}$. Identical shifts have also been seen in the Fourier parameters of single absorption lines so the combination of the lines can be ruled out as an explanation. Also the amplitude/phase diagrams given by MP02 indicate such ``velocity shifts''. 

Close unresolved frequency pairs can cause such a situation if spectroscopically only one component is visible (see Paper I). Looking at the 79 photometrically detected frequencies by Breger et al. (2005)  we see that $\nu_2$, $\nu_3$, $\nu_5$, $\nu_7$\footnote{It could not yet be proven if this mode has a close component or if its amplitude is variable (Breger et al. 2005)} and $\nu_{10}$ indeed have close components, which might distort the amplitude and phase distribution across the line if they are not taken into account in the multi-frequency least-squares solution. 

We calculated new Fourier parameters from a least-squares fit containing 23 frequencies in the solution. All photometrically detected frequencies that are closer than 0.03~\cd~to spectroscopically found frequencies were included. The results show that only the amplitude/phase distribution of $\nu_3$ and $\nu_{10}$ exhibit a significant change. By means of the FPF method $\nu_3$ then can be identified as a radial or dipole mode, as also indicated from photometric mode identification (Breger et al. 1998) and by DD05, whereas $\nu_{10}$ has the best solution at $m=-2$ in a close frequency solution. Unfortunately, the values of $\chi^2_\nu$ are not improved compared to the 14f-solution. An additional problem is that a slight change of the close frequency's value has a large impact on the amplitude/phase values across the line of the mode under consideration.

Both $|f|$ and $\psi_f$ are relatively poorly determined for most modes. The $\chi^2_\nu$ diagrams indicate a wide range of possible $f$ values. This is also true for \alphaw, which is directly dependent on $f$. We have shown that very high S/N data are needed to obtain reliable values of $f$ from the equivalent width variations alone. 
\begin{figure}[!ht]
\centering
  \includegraphics[height=75mm,bb=83 35 523 707,clip,angle=-90]{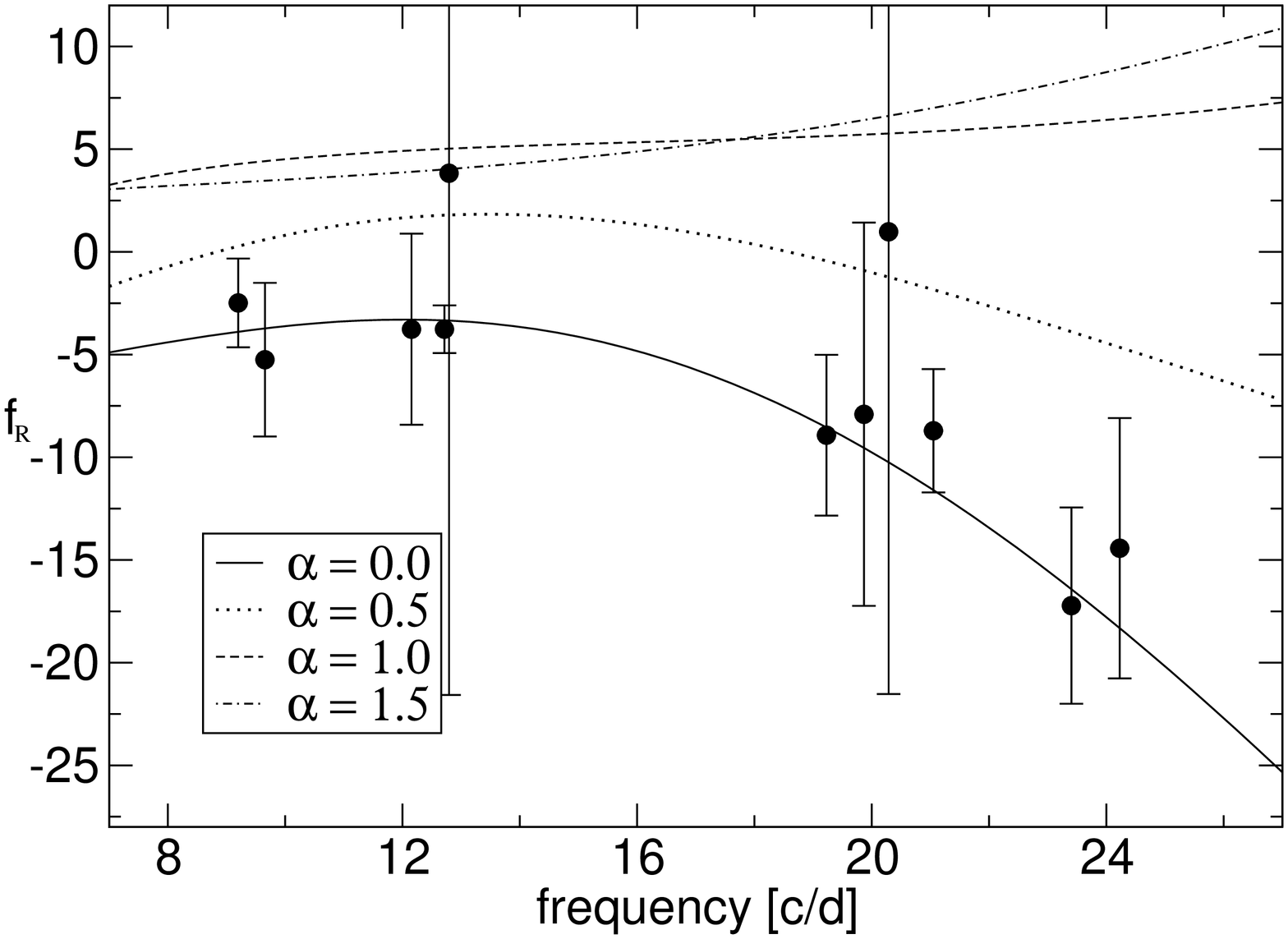}\\
\vspace*{-0cm}
  \includegraphics[height=75mm,bb=83 35 586 707,clip,angle=-90]{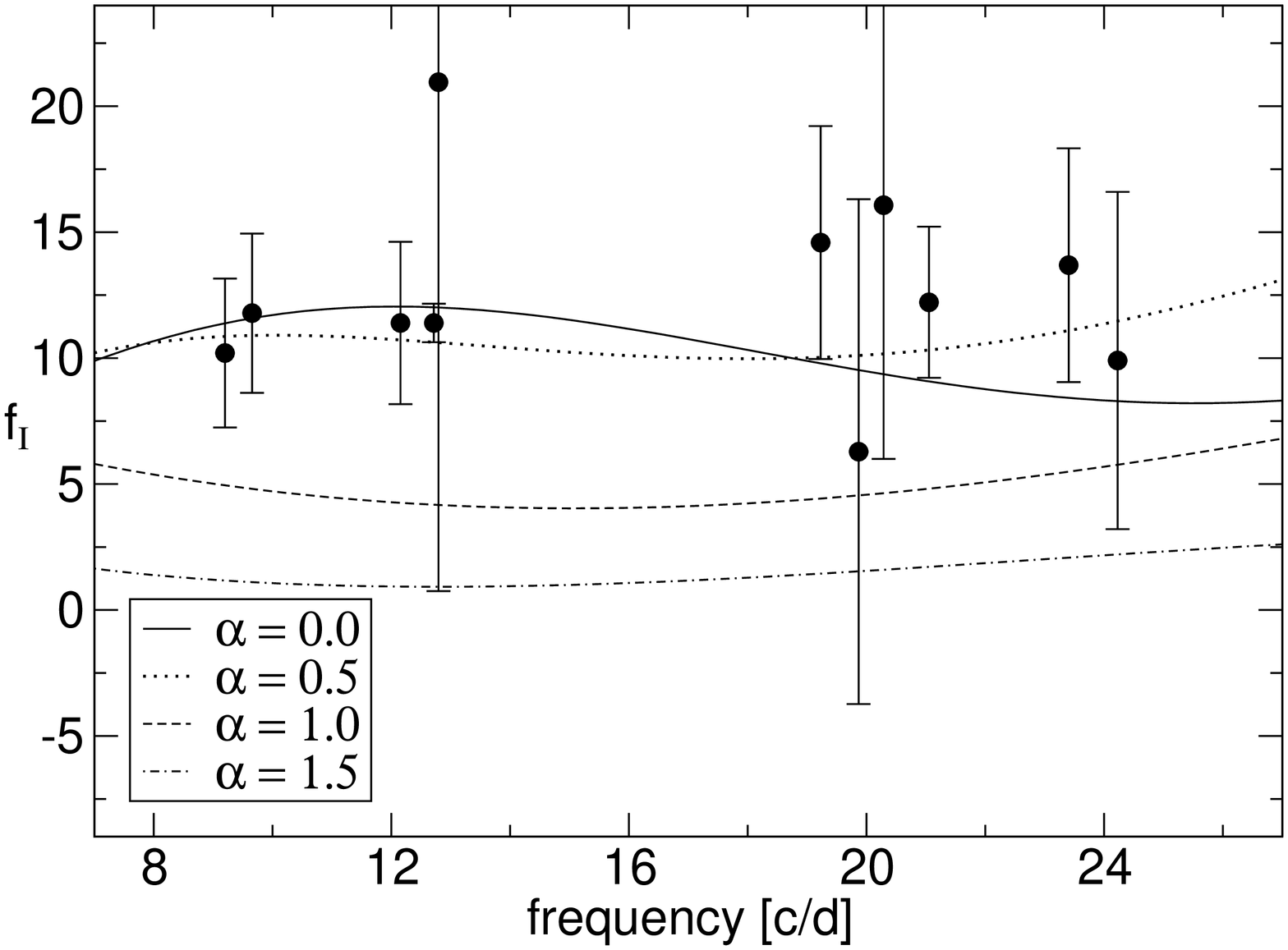}\\
  \caption{Comparison of the empirically derived real (top panel) and imaginary (bottom panel) part of the complex parameter $f$ with theoretical values for different mixing length parameters $\alpha$. The empirical values are shown as black dots with error bars. Each line denotes theoretical values for different values of $\alpha$.}
 \label{fig:empiricalf}
\end{figure}

Taking additional photometric measurements into account significantly improves the quality of the derived $f$. DD05 have shown that it is essential that both spectroscopic and photometric data are obtained in the same year due to the fact that slight year-to-year changes of amplitude and phase may exist. These changes might be due to intrinsic amplitude variability of the star and/or to undetected low-amplitude frequencies. Therefore, we adopted all 79 detected frequencies to a least-squares fit by only taking into account the photometric data from 2002. We adopted the FPF method by simultaneously fitting the Fourier parameters of each mode and its photometric amplitude/phase in $v$ and $y$. Compared to the previous fits, there is no significant difference in the rank of the best identifications of $\ell$ and $m$ for all modes, however, for most of the modes $|f|$, $\psi_f$ and \alphaw~are better constrained now.

It has been shown by Daszynska-Daszkiewicz et al. (2003) that the computed values of $f$ are sensitive to the description of convective transport. In $\delta$~Sct stars like FG Vir two unconnected convective zones associated with the H and HeII ionization zones exist. The energy transport in these convective layers governs the flux variations at the photosphere. A commonly accepted approach to describe the complex convective process is given by the mixing length theory. 
 
In Figure~\ref{fig:empiricalf} we present our empirically derived real and imaginary part of $f$ together with theoretical values for different mixing length parameters $\alpha$.  Note that the sign of our imaginary $f_I$ is inverted compared to DD05 due to the fact that we adopt different conventions of the time dependence $e^{i\omega t}$ of the spherical harmonics. The theoretical values were computed with OPAL opacities (Iglesias \& Rogers 1996) under the assumption of standard Pop I composition ($Y=0.28$, $Z=0.02$). DD05 have shown that the choice of $Z$ has little effect on the computed $f$ values, whereas an increase of $Y$ to 0.33 causes a shift of $f_R$ and $f_I$ in the order of 2. The observations agree well with models having a low mixing length coefficient $\alpha \le 0.5$ meaning that convection is relatively ineffective in FG Vir. This is in good agreement with independently obtained results by DD05.

The results of our mode identification by means of the FPF method are listed in Table~\ref{tab:finallm}. For each mode we have indicated the most probable values of $\ell$ and $m$ from our analysis, together with the independently derived $\ell$-values from DD05 and $m$-values from MP02 for comparison. This table emphasizes the need for independently applied identification techniques. Whereas $\ell$ is much better constrained from the combined photometric/spectroscopic identification technique of DD05, it can not determine the value of $m$. Except for the $m$-values of $\nu_9$ the results of MP02 are coincident with our identification of the azimuthal order.

\begin{table}[!ht]
 \centering
 \caption{Best identification of $\ell$ and $m$ for FG Vir.}
  \begin{tabular}{|c|c|c|c|c|c|}\hline
   name & frequency &\multicolumn{2}{c|}{this work} & DD05 & MP02\\
          & \cd    & $\ell$  &     $m$     &     $\ell$   & $m$\\
      \hline
   $\nu_6$   & 9.199   &1, 2, 3   & 1 &  2 & 1\\
   $\nu_8$   & 9.656   &0, 1, 2& 0 &  2 & 0\\
   $\nu_3$   & 12.154  &0, 1, 2   & 0, 1 &  0 & 0\\
   $\nu_1$   & 12.716  &0, 1      & 0 &  1 & 0\\
   $\nu_{13}$ & 12.794  &2, 3, 4& -2 &  1, 2 & -\\
   $\nu_{9}$  & 19.227  &1, 2   & 1 &  0, 1, 2 & 0\\
   $\nu_7$   & 19.867  &0, 1, 2& 0 &  2, 1 & 0\\
   $\nu_{10}$ & 20.287  &1, 2, 3& -1  &    0, 1 & -\\
   $\nu_{12}$ & 20.834  &2, 3, 4& 1 &  - & -\\
   $\nu_4$   & 21.051  &0, 1, 2& 0 &  1, 0 & 0\\
   $\nu_5$   & 23.403  &2    & 0 &  2, 1 & 0\\
   $\nu_2$   & 24.227  &0, 1   & 0 &  1 & 0\\
\hline
  \end{tabular}
\label{tab:finallm}
\end{table}

\subsection{An improved determination of the inclination}
For the three non-axisymmetric modes $\nu_6$, $\nu_9$ and $\nu_{13}$ we calculated a simultaneous fit with common values of $i$, \vsini, $\sigma$ and $W$. Such an approach is also carried out in the case of the moment method by Briquet \& Aerts (2003). The advantage lies in the simple fact that these four parameters are the same for all these modes. A major drawback of such an approach is the large number of free parameters, which significantly increases the computational time for the determination of the optimum value of the inclination. 

\begin{figure}[!ht]
\centering
  \includegraphics*[height=95mm,clip,angle=-90]{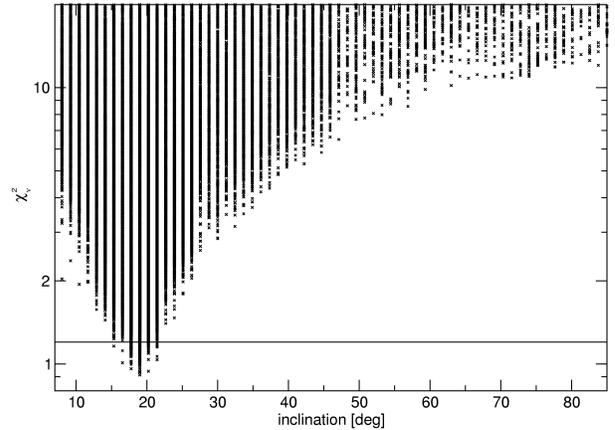}
  \caption{Determination of the inclination of FG Vir from a simultaneous FPF to the three non-axisymmetric modes $\nu_6$, $\nu_9$ and $\nu_{13}$. Every single symbol represents a computed model. The 95~\%-confidence limit of $\chi^2_\nu$ is indicated as solid horizontal line.}
 \label{fig:multimodef3}
\vspace*{3mm}
\end{figure}

We selected the three modes because they can all be fitted within the 95~\%-significance limit. Moreover, they are non-axisymmetric, implying that there is a stronger dependence of $\chi^2_\nu$ with respect to $i$ than for axisymmetric modes (see Paper I).

We applied the multi-mode fit in two steps: with and without taking into account equivalent width variations. We limited the values of $l$ and $m$ to the three best solutions from the mono-mode fits in order to speed up the optimization.

Figure~\ref{fig:multimodef3} shows a clear minimum of the $\chi^2_\nu$ values at $i=19^\circ\pm 5^\circ$. The fit is better for the models that include the equivalent width variations but the derived inclination is the same. Such a small aspect angle was also suspected by MP02, who derived $i=15\pm5$ from a line profile fit to the variations of five modes.

Taking into account \vsini=21.6$\pm$0.3~\kms, we derive an equatorial rotation velocity of $v_{\rm rot}=66\pm16$~\kms. With an assumed radius of $1.58*10^6$~km (Breger et al. 1999) this results in a rotational period of $\Omega=1.74\pm0.43$~d and a rotational frequency of $0.57\pm0.14$~\cd. 

We also see one triplet of rotationally split pulsation modes consisting of $\nu_9$, $\nu_7$ and $\nu_{10}$. From such a triplet it is possible to derive an approximate rotation frequency by considering half of the frequency difference between the $m=-1$ and the $m=+1$ mode. The computed value of 0.53~\cd~is very close to the value independently derived above.

\section{Conclusions}
We carried out a mode identification for the $\delta$~Sct star FG Vir using high-resolution time-series spectra. 15 frequencies could be detected from the radial velocity and pixel-by-pixel intensity variations of a profile made from the combination of four iron lines. 

We have presented the first successful application of the FPF method to line profiles of a $\delta$ Sct star. 
The pulsational geometry of 12 independent frequencies was analyzed by means of this method. Seven frequencies were identified with axisymmetric modes having $\ell$ values between 0 and 2. Five non-axisymmetric modes with $m$ values between -2 and 1 were detected. These modes could be modeled with $\ell$ values as large as $4$. The observed radial velocity amplitudes range between  about 2 and 0.15~\kms. The best models fit the profiles with 0.17~\% relative radius variation for the dominant mode $\nu_1$ and down to a value as small as 0.02~\%. No high degree modes were found mainly due to the low rotational broadening of the line.

By simultaneously fitting amplitudes and phases of velocity and light variations we computed $f$-values for eleven frequencies. A comparison of these empirical $f$-values with theoretical values computed for different mixing length values showed that convection seems to be relatively ineffective in FG Vir.

The good identifications of the azimuthal order, combined with the determination of the inclination angle $i \approx 19^{\circ}$, will help to improve the treatment of rotational effects by seismic models. 

The present fit of the theoretical models to the observations cannot be improved significantly by carrying out additional photometric measurements of FG Vir. Hundreds of unstable modes with $\ell \le 5$ are predicted by theory for the frequency range between 7 and 40~\cd~(Pamyatnykh, priv. comm.). Reducing the observational detection limit from 0.2 to 0.1 mmag would yield many more frequencies but without a knowledge of their pulsational quantum numbers the number of free degrees is increased with every detected frequency. A more promising approach is to start the search for a good stellar pulsation model with the anchor points given by our mode identification.

\begin{acknowledgements}
Thank you to Luciano Mantegazza for the helpful discussions and providing the re-analysis of FG Vir spectra.
WZ was supported by the Austrian Fonds zur F\"orderung der wissenschaftlichen Forschung (Project P17441-N02).
This paper uses observations made at the South African Astronomical Observatory (SAAO).
\end{acknowledgements}

\end{document}